\documentclass[aps,prb,preprint,superscriptaddress]{revtex4-1}

\usepackage{graphicx}
\usepackage{inputenc}
\usepackage{xcolor}
\usepackage{changes}

\newcommand{\um}[0]{\mu\mathrm{m}}

\newcommand{\vkSAW}[0]{\mathbf{k}_\mathrm{SAW}}
\newcommand{\kSAW}[0]{k_\mathrm{SAW}}
\newcommand{\fSAW}[0]{f_\mathrm{SAW}}
\newcommand{\lSAW}[0]{\lambda_\mathrm{SAW}}
\newcommand{\vSAW}[0]{v_\mathrm{SAW}}
\newcommand{\sright}[0]{S_\mathrm{21}}
\newcommand{\sleft}[0]{S_\mathrm{12}}
\newcommand{\epsxx}[0]{\varepsilon_{XX}}
\newcommand{\epsxz}[0]{\varepsilon_{XZ}}
\newcommand{\epszz}[0]{\varepsilon_{ZZ}}

\newcommand{\fSW}[0]{f_\mathrm{SW}}
\newcommand{\vH}[0]{\mathbf{H}}

\newcommand{\Hangle}[0]{\varphi_H}
\newcommand{\damping}[0]{\alpha}
\newcommand{\Pmag}[0]{P_\mathrm{mag}}

\begin{document}

\title{Large non-reciprocal propagation of surface acoustic waves in epitaxial ferromagnetic/semiconductor hybrid structures}
\author{A. Hern\'andez-M\'inguez}
\email{alberto.h.minguez@pdi-berlin.de}
\affiliation{Paul-Drude-Institut f\"ur Festk\"orperelektronik, Leibniz-Institut im Forschungsverbund Berlin e.V., Hausvogteiplatz 5-7, 10117 Berlin, Germany}
\author{F. Maci\`a}
\author{J. M. Hern\`andez}
\affiliation{Dept. of Condensed Matter Physics, University of Barcelona, Mart\'i i Franqu\`es 1, 08028 Barcelona, Spain}
\author{J. Herfort}
\author{P. V. Santos}
\affiliation{Paul-Drude-Institut f\"ur Festk\"orperelektronik, Leibniz-Institut im Forschungsverbund Berlin e.V., Hausvogteiplatz 5-7, 10117 Berlin, Germany}

\date{\today}

\begin{abstract}
Non-reciprocal propagation of sound, that is, the different transmission of acoustic waves traveling along opposite directions, is a challenging requirement for the realization of devices like acoustic isolators and circulators. Here, we demonstrate the efficient non-reciprocal transmission of surface acoustic waves (SAWs) propagating along opposite directions of a GaAs substrate coated with an epitaxial Fe$_3$Si film. The non-reciprocity arises from the acoustic attenuation induced by the magneto-elastic (ME) interaction between the SAW strain field and spin waves in the ferromagnetic film, which depends on the SAW propagation direction and can be controlled via the amplitude and orientation of an external magnetic field. The acoustic transmission non-reciprocity, defined as the difference between the transmitted acoustic power for forward and backward propagation under ME resonance, reaches values of up to 20\%, which are, to our knowledge, the largest non-reciprocity reported for SAWs traveling along a semiconducting piezoelectric substrate covered by a ferromagnetic film. The experimental results are well accounted for by a model for ME interaction, which also shows that non-reciprocity can be further enhanced by optimization of the sample design. These results make Fe$_3$Si/GaAs a promising platform for the realization of efficient non-reciprocal SAW devices.
\end{abstract}

\maketitle

\section{Introduction}\label{sec_Intro}

In the last years, there has been an increasing interest in the realization of non-reciprocal acoustic systems for the efficient manipulation of sound.\cite{Fleury_AcousticsToday11_14_2015} As acoustic propagation is reciprocal in linear systems preserving time reversibility,\cite{Maznev_WaveMotion50_776_2013} it is possible to obtain non-reciprocal transmission by either introducing non-linear effects,\cite{Liang_NatMater9_989_2010, Popa_NatComm5_3398_2014} or by breaking time-inversion symmetry. The last can be achieved by e.g. coupling the acoustic vibrations to a circulating fluid.\cite{Fleury_Science343_516_2014} Another option to induce local non-reciprocity is to take advantage of topologically protected acoustic modes propagating along the boundaries of periodic structures.\cite{Khanikaev_NatComm6_8260_2015, Huber_NatPhys12_621_2016, Chen_PRB98_94302_2018, Fleury_JAcoustSocAm146_719_2019} 
These approaches have been mainly demonstrated for low-frequency sound waves ($\lesssim 1$~MHz) with wavelengths on the order of few millimeters, and the implementation of their working principles in miniaturized acoustic devices working at high frequencies ($\gtrsim 100$~MHz) might result very challenging.

Of special interest for applications are acoustic devices based on surface acoustic waves (SAWs). SAWs are elastic vibrations propagating along the surface of a solid with wavelengths reaching down to the sub-$\mu$m regime and frequencies up to several GHz. Due to their efficient piezoelectric generation and detection, and their low propagation velocities, SAWs have been successfully applied as radio-frequency filters and other kinds of signal processing in on-chip acousto-electric devices.\cite{Campbell___98} In addition to this well-established technology, SAWs are ideally suited for controlling and interfacing elementary excitations in solid-state quantum systems like superconducting circuits,\cite{Gustafsson_Science346_207_2014} defect centers\cite{Golter_PRL116_143602_16, Whiteley_NaturePhysics_2019, Lazic_CommPhys2_113_2019, Iikawa_APL114_171104_2019} and low-dimensional semiconductor structures.\cite{Wiele_PRA58_2680_1998, Couto_NatPhot3_645_2009, Gell_APL93_81115_08, Cerda-Mendez_PRL105_116402_2010, McNeil_N477_439_11, Hermelin_N477_435_11, Lazic_PRB89_85313_14} Therefore, the on-demand controlling of the acoustic propagation direction by means of non-reciprocal devices like SAW isolators and circulators would represent an important step towards the efficient acoustic interfacing between such quantum systems.

Non-reciprocal SAW propagation has been reported in systems like non-magnetic metals\cite{Heil_PRB25_6515_1982} and semiconductor heterostructures,\cite{Zhu_IEEEElectrDevLett38_802_2017} as well as in structures containing ferromagnetic materials.\cite{Lewis_APL20_276_1972, Daniel_JAP48_1732_1977, Weiler_PRL106_117601_11, Dreher_PRB86_134415_12, Sasaki_PRB95_020407_2017} In the first case, non-reciprocity is caused by coupling the lattice strain to the cyclotron motion of free carriers under a strong magnetic field, while in the second case it is related to the non-symmetric transfer of momentum between the acoustic wave and electric currents applied parallel or anti-parallel to the SAW. In the last case, non-reciprocal propagation is induced by coupling the SAW strain field to the magnetization dynamics of a ferromagnet through the magneto-elastic (ME) interaction. As the magnetization precesses only clockwise around its equilibrium direction, the propagation of circularly-polarized acoustic waves along the ferromagnet will depend on the helicity of the strain field relative to the magnetization direction.\cite{Kittel_PhysRev110_836_1958} The excitation of magnetoelastic waves was intensively studied in ferromagnetic insulators like ytrium iron garnet (YIG),\cite{Matthews_APL15_373_1969, Tsutsumi_JAP46_5072_1975, Emtage_PRB13_3063_1976, Komoriya_JAP50_6459_1979, Camley_JAP50_5272_1979} where non-reciprocal SAW attenuation of up to two orders of magnitude was experimentally reported.\cite{Lewis_APL20_276_1972, Daniel_JAP48_1732_1977} However, YIG is a weak magnetoelastic material and, in addition, the non-piezoelectricity of YIG requires the incorporation of strong piezoelectric films like ZnO for the electric excitation of SAWs.\cite{Lewis_APL20_276_1972, Daniel_JAP48_1732_1977, Shimizu_ElectCommJpn63_1_1980, Polzikova_AIPAdv6_056306_2016} 

Another possibility to couple SAWs and magnetization is the deposition of a thin ferromagnetic film at the surface of a piezoelectric substrate. This approach has been demonstrated in LiNbO$_3$ covered by a poly-crystalline cobalt or nickel layer,\cite{Ganguly_JAP47_2696_1976, Davis_APL97_232507_10, Weiler_PRL106_117601_11, Dreher_PRB86_134415_12, Gowtham_JoAP118_233910_15, Labanowski_APL108_022905_2016, Gowtham_PRB94_014436_2016, Sasaki_PRB95_020407_2017, Li_JAP122_043904_2017, Sasaki_PRB99_14418_2019} and in ferromagnetic-semiconductor hybrid structures like GaMnAs/GaAs.\cite{Thevenard_PRB87_144402_2013, Thevenard_PRB90_094401_2014, Thevenard_PRB93_134430_16} Although non-reciprocal propagation has been demonstrated in the first case,\cite{Weiler_PRL106_117601_11, Dreher_PRB86_134415_12, Sasaki_PRB95_020407_2017} the intrinsic structural disorder of poly-crystalline Ni leads to a magnetization response with a relatively large Gilbert damping coefficient $\damping\sim0.05$,\cite{Walowski_JPhysD41_164016_2008} and therefore to wide ME resonances with weak non-reciprocal effects.\cite{Weiler_PRL106_117601_11, Sasaki_PRB95_020407_2017} 

Here, we present an alternative hybrid structure for ME applications consisting of a Fe$_3$Si film grown epitaxially on a GaAs substrate. Fe$_3$Si is a binary Heusler-like ferromagnetic metal which has attracted interest as possible component in magneto-electronic devices.\cite{Herfort_JCrysGrowth278_666_2005, Ionescu_PRB71_94401_2005} As its cubic crystal structure is almost lattice-matched to the GaAs substrate (mismatch $\leq 0.01\%$), it is possible to grow epitaxial films with high interfacial perfection and structural quality,\cite{Herfort_APL83_3912_2003, Thomas_Nanotechnology20_235604_2009, Gusenbauer_PRB83_35319_2011} thus leading to narrow ferromagnetic resonance (FMR) lines\cite{Lenz_Phys.Rev.B72_144411_2005, Wegscheider_PRB84_054461_2011} characterized by damping coefficients as low as $\damping\approx3\times10^{-4}$.\cite{Lenz_pssc3_122_2006} In addition, the shear magneto-elastic coefficient, $b_2$, for thin films of this material has been estimated to be $b_2\approx2-7$~T.\cite{Wegscheider_PRB84_054461_2011} This value is of the same order of magnitude as e.g. crystalline Fe and Ni,\cite{OHandley_230_2000} thus making Fe$_3$Si a promising material for ME applications. Moreover, as SAW control and manipulation of quantum systems has mostly been demonstrated in GaAs-based semiconductor heterostructures,\cite{Couto_NatPhot3_645_2009, Gell_APL93_81115_08, Cerda-Mendez_PRL105_116402_2010, McNeil_N477_439_11, Hermelin_N477_435_11, Lazic_PRB89_85313_14} SAW isolators and circulators based on Fe$_3$Si/GaAs hybrids could provide an efficient integrated acoustic interface between such semiconductor-based quantum devices. Finally, in contrast to GaMnAs/GaAs, where the magnetic properties of GaMnAs require working at cryogenic temperatures,~\cite{Scherbakov_PRL105_117204_2010, Bombeck_PRB85_195324_2012, Thevenard_PRB87_144402_2013, Thevenard_PRB90_094401_2014, Thevenard_PRB93_134430_16} the high Curie temperature (above 800~K)\cite{Nakamura_26_1988} of Fe$_3$Si makes this material suitable for room-temperature applications. 

In this contribution, we demonstrate the non-reciprocal propagation of high-frequency SAWs (3.45~GHz) traveling on a Fe$_3$Si/GaAs hybrid structure. For well-defined orientations of the magnetization in the Fe$_3$Si film, the ME interaction transfers energy from the acoustic into the magnetic system, thus inducing SAW attenuation. The high structural quality of the film leads to very narrow ME-resonances lines (i.e., with full widths at half maximum as narrow as 2~mT). The strength of the ME-induced attenuation depends on the relative angle between magnetization orientation and SAW wave vector. The resulting acoustic transmission non-reciprocity, defined as the difference between the transmitted acoustic power for forward and backward SAW propagation under ME resonance, reaches values up to 20\%. This non-reciprocal behavior is significantly larger than that reported for Ni/LiNbO$_3$ operating at similar frequencies, thus making Fe$_3$Si/GaAs hybrids a promising platform for the realization of non-reciprocal SAW devices like acoustic isolators and circulators in GaAs-based semiconductor heterostructures.

We have organized the manuscript as follows. Section~\ref{sec_Experimental details} describes the fabrication process and the experimental procedure. In Section~\ref{sec_Resul}, we characterize the magnetic and acoustic properties of our sample, and we present the experimental results on the non-reciprocal SAW propagation. In Section~\ref{sec_Disc}, we compare the results with the predictions of the theoretical model and give an outlook. Finally, Section~\ref{sec_Concl} summarizes the main results of the manuscript.

\section{Experimental details}\label{sec_Experimental details}

\begin{figure}
\includegraphics[width=0.5\linewidth]{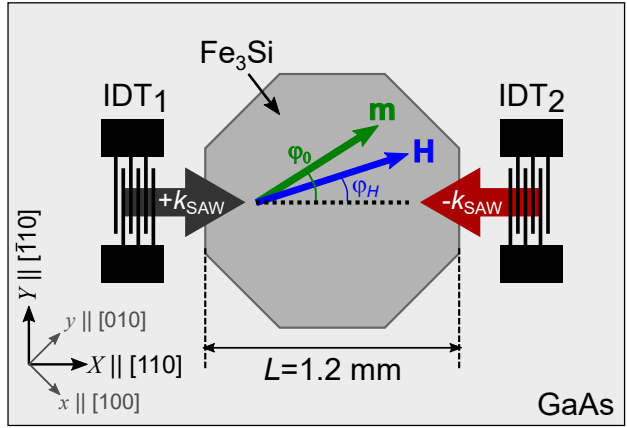}
\caption{Schematics of the magneto-acoustic device. Interdigital transducers (IDTs) at opposite ends of an Fe$_3$Si film launch and detect SAWs with wave vectors $\pm\kSAW$ along the $[110]$ crystallographic direction of the Fe$_3$Si/GaAs hybrid structure. The angles $\Hangle$ and $\varphi_0$ determine the orientation of the external magnetic field, $\vH$, and the equilibrium direction of the magnetization, $\mathbf{m}$, respectively, with respect to the $[110]$ surface direction.}
\label{fig_sample}
\end{figure}

The experiments were performed on a slightly non-stoichiometric Fe$_{3+x}$Si$_{1-x}$ film with $x=0.16$ (corresponding to a Si concentration of 21\%) grown epitaxially on a GaAs(001) substrate by molecular beam epitaxy. Despite the non-stoichiometry, we will further refer to the material as Fe$_3$Si since the structural and magnetic properties of Fe$_{3+x}$Si$_{1-x}$ epitaxial alloys are qualitatively very similar for $-0.07\leq x\leq 0.6$.\cite{Kubaschewski_1982, Herfort_JCSTB22_2073_2004} We fabricated the magneto-acoustic device sketched in Fig.~\ref{fig_sample} as follows. First, a clean c(4$\times$4)-reconstructed GaAs surface was prepared by growing a 500-nm-thick GaAs buffer layer in a dedicated III-V semiconductor growth chamber using conventional growth parameters. The substrate was then transferred in ultra-high-vacuum into an As-free chamber, where the Fe$_3$Si film with a thickness $d=50$~nm was grown by co-deposition from high temperature effusion cells onto the GaAs at 200~$^\circ$C.\cite{Herfort_APL83_3912_2003} Next, we patterned the Fe$_3$Si film by optical lithography and wet chemical etching into octagonal mesas with a distance $L=1.2$~mm between opposite sides. In the final fabrication step, we deposited pairs of interdigital transducers (IDTs) for the generation and detection of SAWs. The IDTs were patterned on the GaAs substrate at the opposite sides of the octagonal Fe$_3$Si mesa by electron beam lithography, metal evaporation and lift-off. Each IDT consists of 180 split-finger\cite{Campbell___98} pairs with a 150~$\um$-wide aperture. The finger periodicity determines the SAW wavelength, which was set to $\lSAW=800$~nm.

Radio-frequency (RF) signals applied to the IDTs excite SAWs propagating with wave vector $\kSAW=2\pi/\lSAW$ parallel ($+\kSAW$) or anti-parallel ($-\kSAW$) to the $[110]$ direction of the Fe$_3$Si/GaAs hybrid structure. The acoustic delay line was characterized by measuring with a vector network analyzer the amplitude of the power transmission coefficient $\sright$ of a SAW traveling from IDT$_1$ to IDT$_2$ ($+\kSAW$), as well as its $\sleft$ counterpart ($-\kSAW$). As the SAW is a Rayleigh mode, the strain tensor consists of three non-zero $\epsxx$, $\epszz$ and $\epsxz$ components,\cite{Rayleigh_PLMSs1-17_4_85} expressed with respect to a rotated reference frame where the $X-$, $Y-$, and $Z-$axes point along the $[110]$, $[\overline{1}10]$ and $[001]$ crystallographic directions, respectively. 

The ME experiments were performed by placing the magneto-acoustic device between the poles of an electromagnet for the application of a static in-plane magnetic field, $\vH$. The sample was mounted on an electrically controlled rotation stage that settles the angle $\Hangle$ between $\vH$ and the [110] surface direction of the Fe$_3$Si/GaAs hybrid structure (see Fig.~\ref{fig_sample}). The angle $\varphi_0$ determines the equilibrium direction of the magnetization, $\mathbf{m}=\mathbf{M}/M_s$ (normalized to the saturation magnetization, $M_s$), defined as the direction that minimizes the magnetic free energy of the Fe$_3S$i film in absence of SAWs (see Supplemental Material). For each values of $\Hangle$ and the magnetic field strength, $H$, we measured both the forward ($\sright$, corresponding to a wave vector $+\kSAW$) and the backward ($\sleft$, corresponding to $-\kSAW$) transmission coefficients of the SAW delay line. Then, we Fourier transformed the frequency-dependent measurements into the time domain to eliminate the electromagnetic cross-talk and analyze the amplitude of the SAW-related transmission peak. All measurements were performed at room temperature.

\section{Results}\label{sec_Resul}

\begin{figure}
\includegraphics[width=0.5\linewidth]{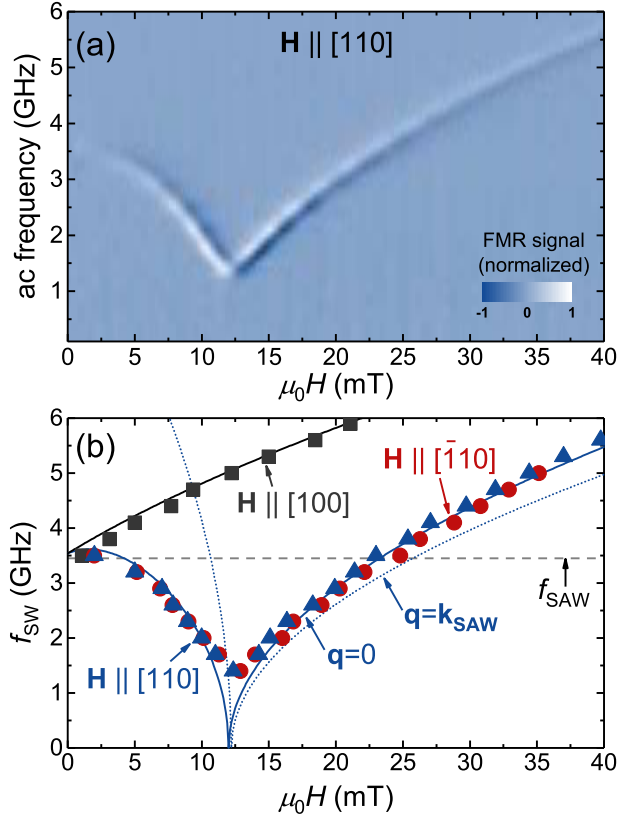}
\caption{(a) Dependence of the ferromagnetic resonance (FMR) signal of the Fe$_3$Si film on the strength of a static magnetic field $\vH$ applied along the $[110]$ direction (horizontal axis), and the frequency of the ac field (vertical axis) applied perpendicular to $\vH$. (b) Spin-wave frequency, $\fSW$, as a function of the magnetic field strength when $\mathbf{H}$ is applied along $[100]$ (black squares), $[\overline{1}10]$ (red circles) and $[110]$ (blue triangles). The solid curves are fittings supposing spin waves with wave vector $\mathbf{q}=0$ (see Supplemental Material). The blue dotted curve is the calculated frequency of spin waves with $\mathbf{q}=\vkSAW$ and $\mathbf{H}$ parallel to $[110]$. The horizontal dashed line indicates the SAW frequency ($\fSAW$) used in our magneto-acoustic device.}
\label{fig_MagProp}
\end{figure}

The magnetic properties of the Fe$_3$Si/GaAs hybrid structure were studied by FMR experiments. The color plot of Fig.~\ref{fig_MagProp}(a) displays the dependence of the FMR signal on the static $\vH$ field applied along $[110]$ (horizontal axis) and on the frequency of an ac magnetic field (vertical axis) applied perpendicularly to $\vH$. The low magnetic field branch ($\mu_0H<10$~mT) of the FMR curve represents the unsaturated state, where $\mathbf{m}$ rotates towards $\vH$ as the magnetic field strength increases. The high magnetic field branch ($\mu_0H>10$~mT) corresponds to the saturated state, where $\mathbf{m}$ and $\vH$ are fully aligned. This behavior indicates that the $[110]$ direction is one of the in-plane hard axes of the magnetization. The narrow FMR lines, with a width of less than 2~mT, attest to the good structural quality and low magnetic damping of the epitaxial material.\cite{Lenz_pssc3_122_2006} Figure~\ref{fig_MagProp}(b) summarizes the dependence of the spin-wave frequency, $\fSW$, on the magnetic field strength when $\vH$ is applied along $[100]$ (black squares), $[\overline{1}10]$ (red circles) and $[110]$ (blue triangles). The solid curves are fittings supposing spin waves with wave vector $\mathbf{q}=0$, that is, the uniform precession mode measured in the FMR experiments (see Supplemental Material for a description of the theoretical model). The overlap of the $[110]$ and $[\overline{1}10]$ curves confirms the negligible in-plane uniaxial anisotropy expected for the Fe concentration, growth conditions and film thickness of this sample.\cite{Lenz_Phys.Rev.B72_144411_2005, Wegscheider_PRB84_054461_2011, Herfort_JAP103_7_2008} Therefore, the in-plane magnetization is only determined by the fourfold cubic crystalline anisotropy with easy axes pointing along $\langle100\rangle$ and hard axes along the $\langle110\rangle$ surface directions.  

\begin{figure}
\includegraphics[width=0.5\linewidth]{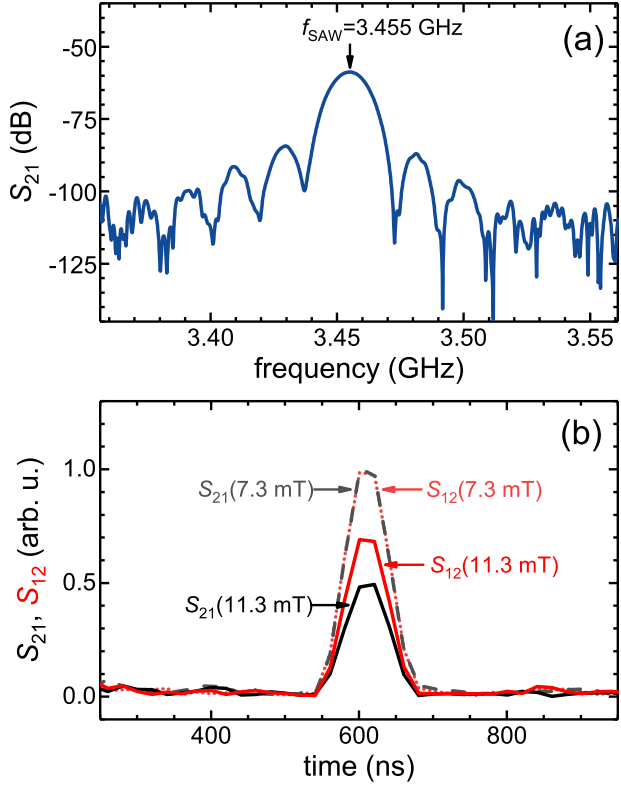}
\caption{(a) Dependence of $\sright$ scattering parameter (corresponding to the RF power transmission coefficient) on the RF frequency applied to IDT$_1$. The spectrum was time-gated to remove the electromagnetic cross-talk. The minimum insertion loss (maximum transmission) occurs at the resonance frequency $\fSAW=3.455$~GHz. (b) Time-resolved $\sright$ and $\sleft$ coefficients measured for $\Hangle=-0.6^\circ$ for two different magnetic field strengths. The peak delays at $\Delta t=610$~ns correspond to the SAW propagation time between the IDTs. The curves are normalized to the magnitude of the $\sright$ transmission peak for $\mu_0H=7.3$~mT.}
\label{fig_DelayLine}
\end{figure}

Figure~\ref{fig_DelayLine}(a) displays the dependence of $\sright$ on the RF frequency applied to IDT$_1$. The measurement was time-gated to remove the electromagnetic cross-talk between IDTs. The transmission spectrum shows a minimum insertion loss of 59~dB (maximum transmission) at the IDT resonant frequency $\fSAW=\vSAW/\lSAW=3.455$~GHz, where $\vSAW$ is the SAW propagation velocity in GaAs. By Fourier transforming the frequency spectrum, we obtained the profile of $\sright$ in the time domain shown in Fig.~\ref{fig_DelayLine}(b) (grey dashed curve). The peak at the time delay $\Delta t\approx610$~ns is attributed to the arrival of the SAW after traveling from IDT$_1$ to IDT$_2$. The value of $\Delta t$ corresponds closely to the acoustic propagation time $\Delta l/\vSAW$ over the center-to-center distance ($\Delta l=1.75$~mm) between the IDTs. The same peak is observed for the time-resolved $\sleft$ coefficient, i.e. for SAW transmission from IDT$_2$ to IDT$_1$ (light red dotted curve).

If the external magnetic field brings the frequency and wave vector of spin waves in the Fe$_3$Si film into resonance with those of the SAW, then the ME interaction will excite spin waves in the ferromagnet for certain angles $\varphi_0$ between $\mathbf{m}$ and the SAW wave vector.\cite{Dreher_PRB86_134415_12} Under these conditions, the ME coupling will convert acoustic into magnetic energy as the SAW propagates along the film, thus resulting in SAW attenuation.\cite{ Weiler_PRL106_117601_11, Dreher_PRB86_134415_12, Thevenard_PRB90_094401_2014, Gowtham_JoAP118_233910_15, Labanowski_APL108_022905_2016, Gowtham_PRB94_014436_2016, Sasaki_PRB95_020407_2017, Li_JAP122_043904_2017} As indicated in Fig.~\ref{fig_MagProp}(b), the frequency of spin waves with wave vector $\mathbf{q}=\vkSAW$ (blue dotted curve) matches the SAW frequency (grey dashed line) for two different strengths of $\vH$ applied along $[110]$. For $\vH$ applied along $[\overline{1}10]$, in contrast, the spin-wave frequency lies always above 10~GHz, and therefore no resonant ME coupling is possible for this magnetic configuration. In the calculations (see Supplemental Material), we have taken into account the wave-vector dependence of the dipolar-dipolar interaction,\cite{Stamps_PRB44_12417_1991, Gowtham_JoAP118_233910_15} but we have neglected the wave-vector dependence of the exchange interaction. This is justified because the spin wave stiffness constant $D=240$~meV\AA$^2$ of Fe$_3$Si (cf. Ref.~\onlinecite{Szymanski_JPCM3_4005_1991}) leads to a frequency shift $\Delta f\propto D\kSAW^2$ of only 36~MHz with respect to the uniform precession mode ($\mathbf{q}=0$) measured in the FMR experiments.

The ME coupling was investigated by measuring $\sright$ and $\sleft$ as a function of $\vH$ applied along the $[\overline{1}10]$ and $[110]$ directions. As expected from our calculations, we did not observe SAW attenuation for $\vH$ parallel to $[\overline{1}10]$. In contrast, the SAW is clearly attenuated for well-defined values of $\vH$ applied along $[110]$. Figure~\ref{fig_DelayLine}(b) compares the time-resolved $\sright$ and $\sleft$ coefficients for two magnetic field strengths applied at an angle $\Hangle=-0.6^\circ$. Away of the ME resonance (for $\mu_0H=7.3$~mT), the intensities of the $\sright$ (gray dashed curve) and $\sleft$ (light red dotted curve) transmission peaks are exactly the same, thus confirming the reciprocal behavior of the SAW delay line. When the magnetic field increases to $\mu_0H=11.3$~mT, the acoustic and magnetic systems enter into resonance leading to a decrease of both $\sright$ and $\sleft$ peaks. Most important, the SAW transmission under ME resonance is clearly different for SAWs propagating with wave vectors $+\kSAW$ and $-\kSAW$. For $\sright$ (solid black curve), the ME coupling reduces the transmitted SAW power to 50\% of the corresponding out-of-resonance value, while for $\sleft$ (solid red curve) the transmission of the SAW peak is still 70\% of the out-of-resonance one. The difference between these values yields a transmission non-reciprocity of 20\% for SAWs traveling along opposite directions.

\begin{figure}[t]
\includegraphics[width=0.7\linewidth]{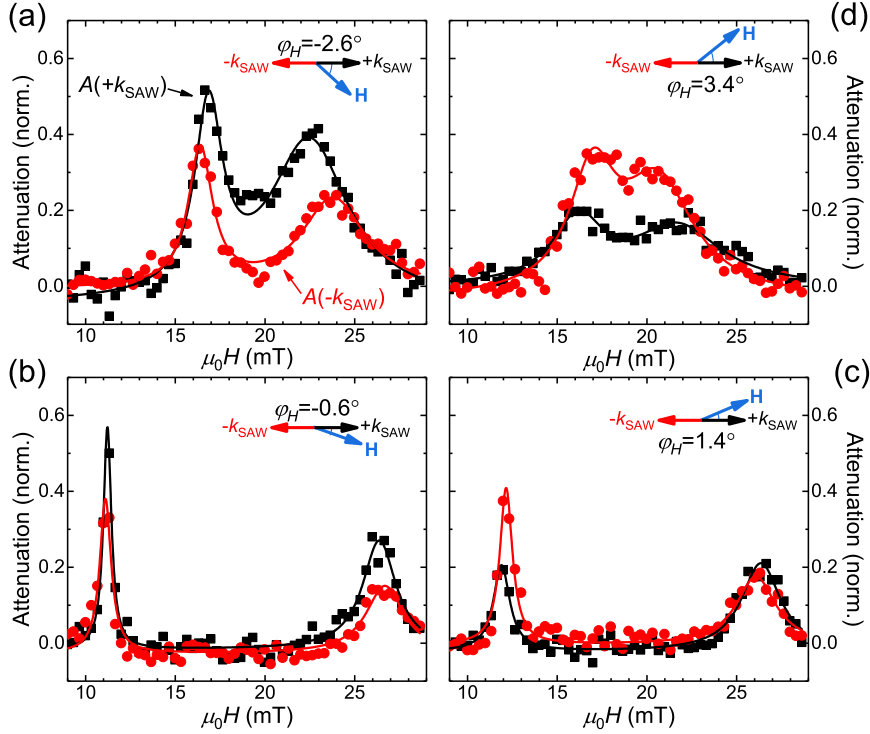}
\caption{Dependence of the ME-induced SAW attenuation, $A$, on the magnetic field amplitude, $H$, for SAWs with $+\kSAW$ (black squares) and $-\kSAW$ (red circles). The panels show measurements taken for magnetic field angles: (a) $\Hangle=-2.6^\circ$, (b) $\Hangle=-0.6^\circ$, (c) $\Hangle=1.4^\circ$, and (d) $\Hangle=3.4^\circ$. The solid curves are Lorentzian fits to the ME resonance lines.}
\label{fig_MEW_Hscans}
\end{figure}

To get further insight into the non-reciprocal behavior, we have measured $\sright$ and $\sleft$ for a range of $H$ and $\Hangle$ values to determine the SAW attenuation $A=1-T$. Here, $T$ represents the area of the SAW transmission peak in the time-domain spectrum, normalized to the corresponding area for $H$ away from the ME resonance. Figure~\ref{fig_MEW_Hscans} shows the dependence of $A(\pm\kSAW)$ on $H$ measured at four angles $\Hangle$. Each trace shows two ME resonances at two field values. For the resonance at the low magnetic field, $\mathbf{m}$ is still rotating towards $\vH$, while $\mathbf{m}$ and $\vH$ are fully aligned for the resonance at high magnetic field. The magnetic field distance between the two resonant fields is maximum for $\Hangle\approx0$, and their position agrees well with the values predicted by the crossing between spin-wave and SAW frequencies in Fig.~\ref{fig_MagProp}(b). As $\vH$ rotates away from the $[110]$ direction, the resonances move towards each other until they merge at $|\Hangle|\approx4^\circ$. For larger values of $\Hangle$, the frequency of the spin waves lies above the SAW frequency, and no ME resonances can be excited. For all four orientations of $\Hangle$ in Fig.~\ref{fig_MEW_Hscans}, $A(+\kSAW)$ (black squares) is clearly different from $A(-\kSAW)$ (red circles) at the ME resonance, reaching a non-reciprocal attenuation efficiency $\Delta A=A(+\kSAW)-A(-\kSAW)\approx\pm20$\%. Moreover, the sign of $\Delta A$ depends on the sign of the magnetic field angle: for $\Hangle<0$, $\Delta A>0$, cf. Figs.~\ref{fig_MEW_Hscans}(a) and \ref{fig_MEW_Hscans}(b), while for $\Hangle>0$ the situation reverses and $\Delta A<0$, cf. Figs.~\ref{fig_MEW_Hscans}(c) and \ref{fig_MEW_Hscans}(d).

\begin{figure}
\includegraphics[width=0.9\linewidth]{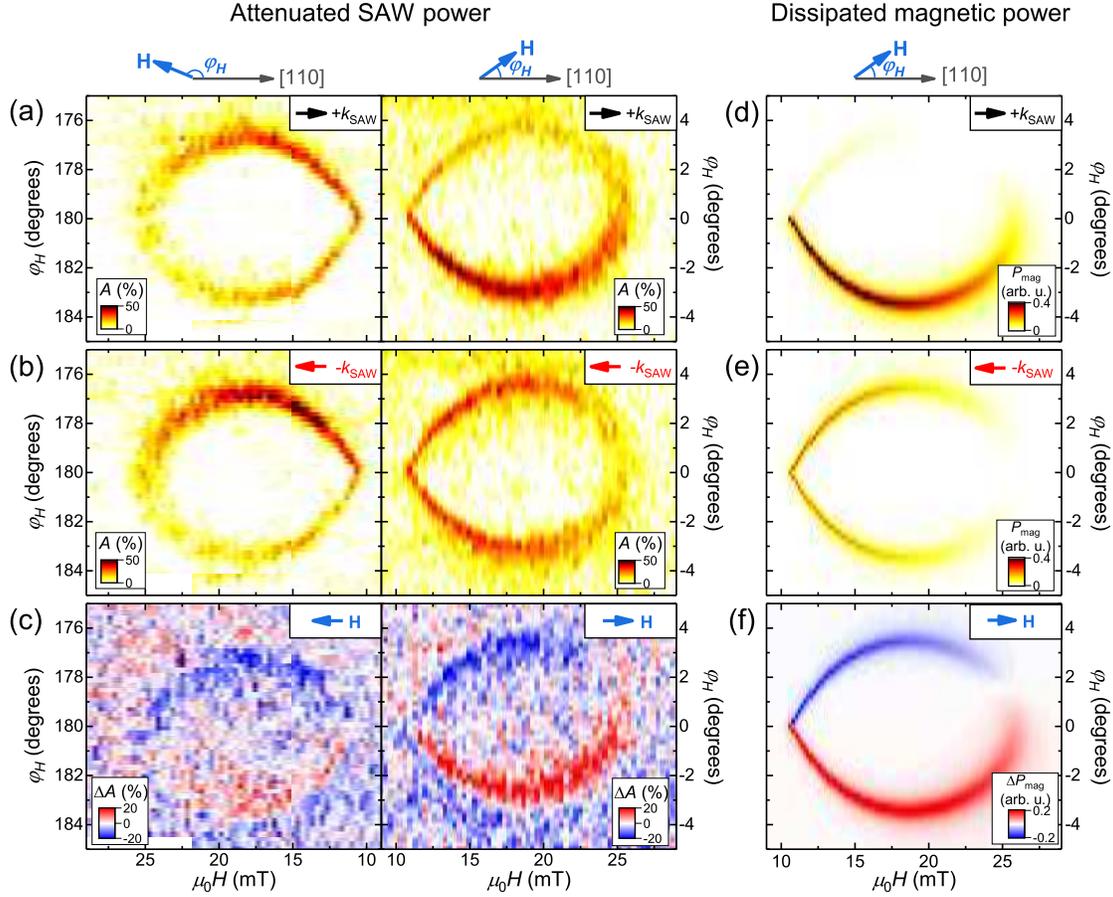}
\caption{ME-induced attenuation maps as a function of magnetic field strength, $H$, and angle, $\Hangle$, for SAWs propagating with wave vectors (a) $+\kSAW$ and (b) $-\kSAW$. (c) SAW attenuation difference, $\Delta A$, calculated from the data in panels (a) and (b). (d) Simulation of the magnetic power dissipated by the Fe$_3$Si film, $\Pmag$, as a function of $H$ and $\Hangle$ for SAW with wave vector $+\kSAW$. (e) Same as (d), but for SAW with wave vector $-\kSAW$. (f) Difference $\Delta\Pmag$ between panels (d) and (e).}
\label{fig_MEW_Summary}
\end{figure}

The two-dimensional color plots of Fig.~\ref{fig_MEW_Summary}(a) and \ref{fig_MEW_Summary}(b) summarize the dependence of the SAW attenuation on the strength $H$ and angular direction $\Hangle$ (cf. sketch in the upper part of the figure) of the magnetic field. The resonant ME interaction (and, consequently, SAW attenuation) takes place only on a $\Hangle\times H$ lobe defined by a very narrow range of angles $\Hangle$ around $0^\circ$ and $180^\circ$. Figure~\ref{fig_MEW_Summary}(a) shows experimental data recorded for $\vH$ directions quasi-parallel (right panel, $-5^\circ<\Hangle<5^\circ)$ and quasi-antiparallel (left panel, $175^\circ<\Hangle<185^\circ$) to $+\kSAW$. Figure~\ref{fig_MEW_Summary}(b) displays the corresponding data for SAW with wave vector $-\kSAW$. For all configurations, the attenuation changes as the magnetic field rotates away from the $[110]$ axis. For the quasi-parallel configurations (right panel of Fig.~\ref{fig_MEW_Summary}(a) and left panel of Fig.~\ref{fig_MEW_Summary}(b)), the magnitude of the attenuation is strongly angular dependent with clearly different values in the upper and lower side of the $\Hangle\times H$ lobes. In the quasi-antiparallel case, in contrast, the angular dependence is less pronounced. As it will be discussed in the next section, this weaker dependence arises from small deviations of the SAW wave vector from the $[110]$ axis.

Finally, we show in Fig.~\ref{fig_MEW_Summary}(c) the non-reciprocal attenuation efficiency, $\Delta A$, determined from the difference between the data in the corresponding panels of Fig.~\ref{fig_MEW_Summary}(a) and \ref{fig_MEW_Summary}(b). Both panels of this figure show different signs of $\Delta A$ in the upper and lower sides of the $\Hangle\times H$ lobes. As in Fig.~\ref{fig_MEW_Hscans}, $\Delta A$ in Fig.~\ref{fig_MEW_Summary}(c) reaches values as large as $\pm20$\%. Taking this value together with the length $L$ of the ferromagnetic film, we estimate a non-reciprocal attenuation rate, $\eta\approx\Delta A/L\approx16$~\%/mm. For comparison, we have also estimated $\eta$ from the attenuation values reported in the Ni/LiNbO$_3$ hybrid structures working at 2.24~GHz SAW frequency,\cite{Weiler_PRL106_117601_11, Sasaki_PRB95_020407_2017} obtaining $\eta\sim1.6\pm0.9$~\%/mm. The latter is one order of magnitude lower than the ones obtained in the Fe$_3$Si/GaAs structures reported here.

\section{Discussion}\label{sec_Disc}

To theoretically analyze the experimental results, we have taken into account that, according to energy conservation, the attenuated SAW power must equal the power dissipated by the spin waves.\cite{Dreher_PRB86_134415_12, Gowtham_PRB94_014436_2016} Therefore, as a first approximation to the problem, we have just estimated the response of the magnetization to the SAW-induced ME field, and compared the power dissipated by the spin waves to the SAW attenuation profiles of Fig.~\ref{fig_MEW_Summary}. The magnetization dynamics is described by the Landau-Lifshitz-Gilbert (LLG) equation:\cite{Gilbert_PR100_1243_1955, Gilbert_IEEETransMagn40_3443_2004}

\begin{equation}\label{eq_LLG}
\dot{\mathbf{m}}=-\gamma\mathbf{m}\times\mu_0\vH_\mathrm{eff}+\damping\mathbf{m}\times\dot{\mathbf{m}},
\end{equation}

\noindent where $\mu_0$ is the vacuum permeability, $\gamma$ is the gyromagnetic ratio, $\damping$ is the Gilbert damping constant, and the dot denotes the time derivative. The temporal evolution of $\mathbf{m}$ depends on the effective magnetic field $\mu_0\vH_\mathrm{eff}=-\overrightarrow{\nabla}_{\mathbf{m}}F$. Here, $F$ is the free magnetic energy normalized to the saturation magnetization. In addition to the Zeeman, shape and crystalline anisotropiy energies (see Supplemental Material), it includes the magneto-elastic energy, $F_{me}$, that couples the oscillating strain of the SAW to the magnetization. In a material with cubic symmetry, the lowest order contributions to $F_{me}$ can be expressed as:\cite{OHandley_230_2000}

\begin{eqnarray}\label{eq_Fmecubic}
F_{me}&=&b_1[\varepsilon_{xx}m_x^2+\varepsilon_{yy}m_y^2+\varepsilon_{zz}m_z^2]\nonumber\\&&+2b_2[\varepsilon_{xy}m_xm_y+\varepsilon_{xz}m_xm_z+\varepsilon_{yz}m_ym_z],
\end{eqnarray}

\noindent where $b_1$ and $b_2$ represent the longitudinal and shear ME coefficients, respectively, while $\varepsilon_{ij}$  and $m_i$ are the projections of the strain and  magnetization components onto the $\hat{\mathbf{x}}\parallel[100]$, $\hat{\mathbf{y}}\parallel[010]$ and $\hat{\mathbf{z}}\parallel[001]$ directions of the cubic lattice. As mentioned above, it is convenient to rewrite Eq.~\ref{eq_Fmecubic} as a function of the three non-zero strain components $\epsxx$, $\epszz$ and $\epsxz$ of the SAW expressed in the rotated $XYZ$ reference frame, see Fig.~\ref{fig_sample}. We will also describe $\mathbf{m}$ in spherical coordinates with azimuthal and polar angles, $\varphi$ and $\theta$, expressed with respect to $[110]$ and $[001]$, respectively. A detailed derivation of the equations is presented in the Supplemental Material. For small angular deviations $\delta\varphi$ and $\delta\theta$ of $\mathbf{m}$ with respect to its equilibrium direction, the effective ME field driving the magnetization precession, $\mu_0\mathbf{h}=-\overrightarrow{\nabla}_\mathbf{m}F_{me}$, consists of the following in-plane and out-of-plane components perpendicular to $\mathbf{m}$,  $h_\varphi$ and $h_\theta$, respectively:

\begin{eqnarray}
\mu_0h_\varphi&=&2b_2\sin(\varphi_0)\cos(\varphi_0)\epsxx,\label{eq_hvarphi}\\
\mu_0h_\theta&=&2b_2\cos(\varphi_0)\epsxz.\label{eq_htheta}
\end{eqnarray}

\begin{figure}
\includegraphics[width=0.5\linewidth]{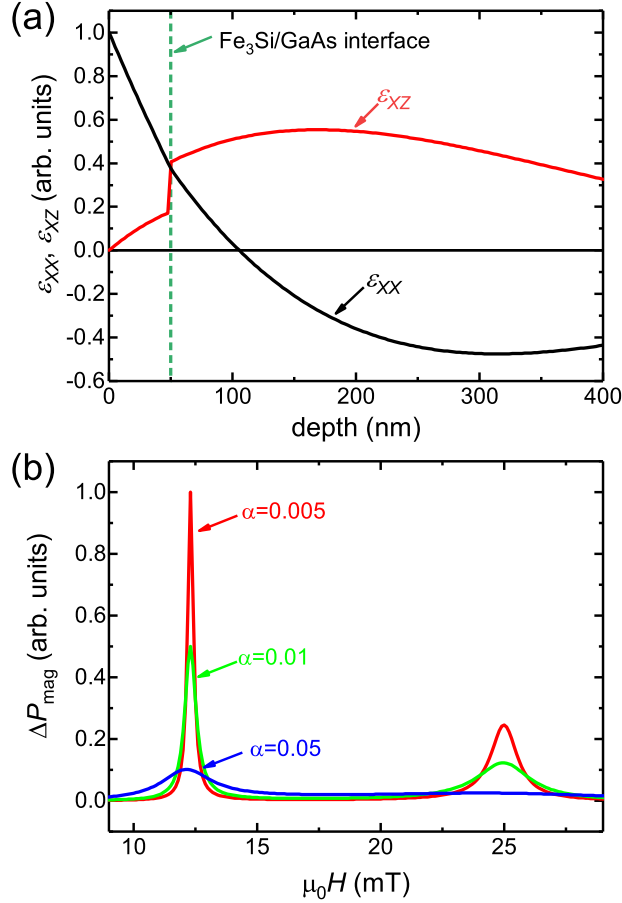}
\caption{(a) Dependence of the longitudinal and shear strain amplitudes, $\epsxx$ (black curve) and $\epsxz$ (red curve), respectively, on sample depth for a SAW with 800~nm wavelength. The curves are normalized with respect to the amplitude of $\epsxx$ at the top surface. The vertical dashed line indicates the position of the Fe$_3$Si/GaAs interface. (b) Difference in dissipated magnetic power, $\Delta\Pmag$, for the parallel and anti-parallel configurations of $H$ and $\kSAW$, as a function of the Gilbert damping coefficient $\damping$. The data were calculated for $\fSAW=3.45$~GHz, $\Hangle=-1.5^\circ$, $\epsxz=0.5\epsxx$ and $\beta=0$, and are normalized with respect to the maximum value for $\damping=0.005$.}
\label{fig_SAWprofile}
\end{figure}

\noindent The in-plane component $h_\varphi$ is proportional to the longitudinal strain of the SAW, $\epsxx$, while the out-of-plane component $h_\theta$ is proportional to the shear strain, $\epsxz$. Both compoments depend on the in-plane equilibrium direction of $\mathbf{m}$ through $\varphi_0$ (see Fig.~\ref{fig_sample}). In the absence of an external magnetic field, $\mathbf{m}$ points along one of the $\langle100\rangle$ easy axes (i.e., $\varphi_0=\pm45^\circ$). When $\vH$ is applied along the $[\overline{1}10]$ direction, $\mathbf{m}$ rotates toward $\varphi_0\approx 90^\circ$, so that both $h_\varphi$ and $h_\theta$ vanish. This, together with the large spin wave frequencies expected for this magnetic configuration, explain the absence of ME coupling for this orientation of $\vH$. In contrast, when $\vH$ points along $[110]$, $\mathbf{m}$ rotates toward $\varphi_0\approx0$. Here, $h_\varphi$ also tends to zero, but now $h_\theta$ approaches its maximum value, provided the SAW strain component $\epsxz$ is not zero. As $\epsxz$ vanishes at a free surface, $h_\theta$ is usually very small for thin ferromagnetic films and/or large SAW wavelengths, and can normally be neglected.\cite{Thevenard_PRB87_144402_2013, Kuszewski_JPhysCM30_244003_2018, Kuszewski_PRApp10_034036_2018} For thicker films and/or shorter SAW wavelengths, however, $\epsxz$ becomes relevant, and the contribution of $h_\theta$ to the magnetization dynamics must be taken into account. To visualize this behavior, we display in Fig.~\ref{fig_SAWprofile}(a) the amplitude of $\epsxx$ and $\epsxz$ as a function of sample depth calculated for a $\lSAW=800$~nm SAW. The calculations of the SAW strain were carried out using an elastic model that takes into account the elastic properties of Fe$_3$Si\cite{Koetter_MaterSciEngA114_29_1989} and GaAs\cite{Auld90a}. While $\epsxx$ decreases from its maximum value as it crosses the film, $\epsxz$ increases from zero and reaches a value $\epsxz\approx0.5\epsxx$ at the Fe$_3$Si/GaAs interface (marked as a vertical dashed line in Fig.~\ref{fig_SAWprofile}(a)).

The observed non-reciprocal SAW attenuation can therefore be understood as an interplay between the $h_\varphi$ and $h_\theta$ components.\cite{Dreher_PRB86_134415_12, Sasaki_PRB95_020407_2017} Due to the $\pi/2$ time phase shift between $\epsxx$ and $\epsxz$,\cite{Rayleigh_PLMSs1-17_4_85} the ME driving field $\mathbf{h}=h_\varphi\hat{\varphi}+ih_\theta\hat{\theta}$ is, in general, elliptically polarized. Its helicity depends on the ratio $\epsxz/\epsxx$ and changes signs when one reverses the SAW propagation direction. As the magnetization precession described by the LLG equation has also a well-defined helicity, the strength of the ME coupling will depend on the helicity of $\mathbf{h}$.\cite{Stancil_2009} For example, for a SAW propagating with $+\kSAW$, $\epsxx\propto\cos\omega t$ and $\epsxz\propto\sin\omega t$. If $\vH$ rotates the magnetization towards $0<\varphi_0<45^\circ$, then $\mathbf{h}$ and $\mathbf{m}$ will precess with opposite  helicities and the coupling of the SAW to the magnetization dynamics will be weak, see right panel in Fig.~\ref{fig_MEW_Summary}(a). However, if $\vH$ orients $\mathbf{m}$ towards $-45^\circ<\varphi_0<0$, the sign of $h_\varphi$ will reverse and now both $\mathbf{h}$ and $\mathbf{m}$ will precess with the same helicity. Under these conditions, the ME coupling will be strong, thus inducing a larger SAW attenuation. When $\vH$ reverses and the magnetization points against the SAW wave vector, $h_\theta$ changes sign and inverts the helicity dependence of $\mathbf{h}$ on $\varphi_0$. The same happens if the magnetization stays along $[110]$, but the SAW propagates with $-\kSAW$, causing $\epsxz\propto\sin\omega t$ to be replaced by $\epsxz\propto-\sin\omega t$. Finally, when both magnetization and SAW wave vector are reversed, $h_\theta$ remains positive and the original dependence on the relative angle between $\vH$ and SAW wave vector is recovered, see the left panel in Fig.~\ref{fig_MEW_Summary}(b).

According to the previous discussion, the SAW attenuation profiles for quasi-parallel and quasi-antiparallel configurations of $\vH$ and SAW wave vector in Fig.~\ref{fig_MEW_Summary}(a) and \ref{fig_MEW_Summary}(b) should be mirror images of each other. However, as already mentioned in the previous section, this is not exactly the case for the experimental data. We attribute this discrepancy to a small, unintentional misalignment angle $\beta$ between the SAW wave vector and the $[110]$ direction, probably caused during the patterning of the IDTs. This misalignment breaks the symmetry of $h_\varphi$ and $h_\theta$ by introducing additional terms that depend on $\beta$ and $b_1$ (see Supplemental Material). As a consequence, the non-reciprocity with respect to $\varphi_0$ is enhanced in the quasi-parallel configuration, but partially compensated in the quasi-antiparallel configuration. To confirm this assumption, we have estimated the dependence on $H$ and $\Hangle$ of the power dissipated by the spin waves and compared it to the profiles in Fig.~\ref{fig_MEW_Summary}(a) and (b). The dissipated magnetic power, $\Pmag$, can be calculated as:\cite{Dreher_PRB86_134415_12}

\begin{equation}\label{eq_power}
\Pmag=-\mathrm{Im}\left[\mu_0\frac{\omega}{2}\int_{V_0}\left(\mathbf{h}^*\overline{\chi}\mathbf{h}\right)\mathrm{d}V\right],
\end{equation}

\noindent where $\overline{\chi}$ is the Polder susceptibility tensor describing the response of the magnetization to $\mathbf{h}$, and $V_0$ is the volume of the ferromagnetic film traversed by the SAW. To obtain $\overline{\chi}$, we have solved the linearized LLG equation supposing harmonic solutions for $\epsxx$, $\epsxz$, $\delta\theta$ and $\delta\varphi$. A detailed derivation is presented in the Supplemental Material. Although an accurate calculation of $\Pmag$ requires the integration of $\mathbf{h}^*\overline{\chi}\mathbf{h}$ taking into account the variation of $\epsxx$ and $\epsxz$ across the depth of the Fe$_3$Si film, we have just estimated $\mathbf{h}^*\overline{\chi}\mathbf{h}$ close to the Fe$_3$Si/GaAs interface. This depth yields the largest $\epsxz/\epsxx$ ratio, see Fig.~\ref{fig_SAWprofile}(a), where one expects the strongest non-reciprocal effect. In the simulation, we have used the values of the gyromagnetic ratio, shape and cubic anisotropies obtained from the fitting of the FMR curves in Fig.~\ref{fig_MagProp}. The width of the ME resonance depends on the damping coefficient, which was set to $\damping=8\times10^{-3}$. The degree of non-reciprocity depends on the values of $b_1$, $b_2$, $\epsxx$, $\epsxz$ and $\beta$, which we have set to  $b_1=6$~T, $b_2=2$~T, $\epsxx=10^{-4}$, $\epsxz=0.5\times10^{-4}$ and $\beta=1^\circ$. Figures~\ref{fig_MEW_Summary}(d) and \ref{fig_MEW_Summary}(e) display $\Pmag$ for $\vH$ and SAW wave vector in the quasi-parallel and quasi-antiparallel configurations, respectively, while the difference $\Delta\Pmag$ between both configurations is shown in Fig.~\ref{fig_MEW_Summary}(f). The theoretical model reflects qualitatively well the position of the ME resonances, their amplitude dependence on $H$ and $\Hangle$ and the non-reciprocal behavior.

To conclude this section, we briefly discuss the reasons for the superior non-reciprocal SAW propagation of the Fe$_3$Si/GaAs hybrid structure compared to the poly-crystalline nickel on LiNbO$_3$. The first is the high degree of structural quality of our epitaxial Fe$_3$Si film, which ensures the low Gilbert damping coefficient required for the observation of the strong non-reciprocal effects. To illustrate this point, we have estimated $\Delta\Pmag$ as a function of $H$ for a fixed value of $\Hangle$ and several values of $\damping$. The results are displayed in Fig.~\ref{fig_SAWprofile}(b), normalized with respect to the maximum value for $\damping=0.005$. We observe that $\Delta\Pmag$ is inversely proportional to $\damping$, and becomes ten times smaller when $\damping$ increases from 0.005 to 0.05. This is in good agreement with the difference in the calculated values of $\eta$ for the Fe$_3$Si/GaAs and Ni/LiNbO$_3$ systems.

The second reason is the short SAW wavelength used in our experiment (800~nm), which is about two times smaller than in the Ni/LiNbO$_3$ (1.5~$\mu$m) structures. This implies a larger $\epsxz/\epsxx$ ratio at the ferromagnetic film and, therefore, a larger degree of elliptical polarization of the effective ME field driving the magnetization precession. Following this approach, it should be possible to further improve the sample design to get SAW isolators with non-reciprocal transmission exceeding 10~dB. This will happen when the SAW-induced ME field is fully circularly polarized, that is, when $|h_\varphi|=|h_\theta|$. According to Eqs.~\ref{eq_hvarphi} and \ref{eq_htheta}, this condition fulfills when $\epsxz/\epsxx=\sin(\varphi_0)$. For instance, when $\epsxz/\epsxx=0.5$, the maximum non-reciprocity will happen if ME resonance takes place for $\varphi_0=30^\circ$. It should be possible to get close to these conditions by an accurate selection of Fe$_3$Si film thickness and SAW frequency.

Finally, an alternative approach to increase the degree of non-reciprocity has recently been proposed.\cite{Verba_PRApp9_064014_2018, Verba_PRApp12_54061_2019} It consists in modifying the spin-wave frequency dispersion curves, so that the frequency crossing points for SAWs and spin waves propagating along $-\kSAW$ differ from the ones expected in the $+\kSAW$ case. This can be done e.g. by covering the Fe$_3$Si film with a heavy metal capping layer,\cite{Verba_PRApp9_064014_2018} or by fabricating structures consisting of two ferromagnetic films close to each other with opposite magnetization directions.\cite{Verba_PRApp12_54061_2019} As epitaxial trilayers consisting of two Fe$_3$Si films separated either by a thin metal\cite{Jenichen_SST30_114005_2015} or semiconductor spacer\cite{Gaucher_APL110_102103_2017, Gaucher_SST33_104005_2018} have already been reported, Fe$_3$Si/GaAs hybrid structures could be a promising system for the experimental implementation of this mechanism.

\section{Conclusions}\label{sec_Concl}

In this contribution, we have demonstrated non-reciprocal propagation of SAWs along a GaAs substrate covered with an epitaxial Fe$_3$Si film. For well-defined values of the external magnetic field, the magneto-elastic coupling transfers energy from the acoustic into the magnetic system, thus inducing attenuation of the SAW. The strength of the SAW attenuation depends on the relative orientation between magnetization and SAW wave vector, and leads to attenuation differences of up to 20\% for SAWs propagating along opposite directions. We attribute the non-reciprocal behavior to the dependence of the magnetization dynamics on the helicity of the elliptically polarized magneto-elastic field associated to the SAW. Our simulations confirm these results and show that non-zero longitudinal and shear strain at the ferromagnetic film, as well as low magnetic damping, are critical to achieve large non-reciprocal effects. Due to the combination of large magnetostriction, low magnetic damping, and the possibility of growing multilayers, epitaxial Fe$_3$Si/GaAs hybrids are promising systems for the realization of non-reciprocal acoustic devices in GaAs-based semiconductor heterostructures.

\section*{Acknowledgements}
The authors would like to thank Manfred Ramsteiner for a critical reading of the manuscript, as well as H.-P. Sch\"onherr, S. Rauwerdink and S. Meister for technical support in the preparation of the samples.


\begin{thebibliography}{78}%
\makeatletter
\providecommand \@ifxundefined [1]{%
 \@ifx{#1\undefined}
}%
\providecommand \@ifnum [1]{%
 \ifnum #1\expandafter \@firstoftwo
 \else \expandafter \@secondoftwo
 \fi
}%
\providecommand \@ifx [1]{%
 \ifx #1\expandafter \@firstoftwo
 \else \expandafter \@secondoftwo
 \fi
}%
\providecommand \natexlab [1]{#1}%
\providecommand \enquote  [1]{``#1''}%
\providecommand \bibnamefont  [1]{#1}%
\providecommand \bibfnamefont [1]{#1}%
\providecommand \citenamefont [1]{#1}%
\providecommand \href@noop [0]{\@secondoftwo}%
\providecommand \href [0]{\begingroup \@sanitize@url \@href}%
\providecommand \@href[1]{\@@startlink{#1}\@@href}%
\providecommand \@@href[1]{\endgroup#1\@@endlink}%
\providecommand \@sanitize@url [0]{\catcode `\\12\catcode `\$12\catcode
  `\&12\catcode `\#12\catcode `\^12\catcode `\_12\catcode `\%12\relax}%
\providecommand \@@startlink[1]{}%
\providecommand \@@endlink[0]{}%
\providecommand \url  [0]{\begingroup\@sanitize@url \@url }%
\providecommand \@url [1]{\endgroup\@href {#1}{\urlprefix }}%
\providecommand \urlprefix  [0]{URL }%
\providecommand \Eprint [0]{\href }%
\providecommand \doibase [0]{http://dx.doi.org/}%
\providecommand \selectlanguage [0]{\@gobble}%
\providecommand \bibinfo  [0]{\@secondoftwo}%
\providecommand \bibfield  [0]{\@secondoftwo}%
\providecommand \translation [1]{[#1]}%
\providecommand \BibitemOpen [0]{}%
\providecommand \bibitemStop [0]{}%
\providecommand \bibitemNoStop [0]{.\EOS\space}%
\providecommand \EOS [0]{\spacefactor3000\relax}%
\providecommand \BibitemShut  [1]{\csname bibitem#1\endcsname}%
\let\auto@bib@innerbib\@empty
\bibitem [{\citenamefont {Fleury}\ \emph {et~al.}(2015)\citenamefont {Fleury},
  \citenamefont {Sounas}, \citenamefont {Haberman},\ and\ \citenamefont
  {Alù}}]{Fleury_AcousticsToday11_14_2015}%
  \BibitemOpen
  \bibfield  {author} {\bibinfo {author} {\bibfnamefont {R.}~\bibnamefont
  {Fleury}}, \bibinfo {author} {\bibfnamefont {D.}~\bibnamefont {Sounas}},
  \bibinfo {author} {\bibfnamefont {M.~R.}\ \bibnamefont {Haberman}}, \ and\
  \bibinfo {author} {\bibfnamefont {A.}~\bibnamefont {Alù}},\ }\href@noop {}
  {\bibfield  {journal} {\bibinfo  {journal} {Acoustics Today}\ }\textbf
  {\bibinfo {volume} {11}},\ \bibinfo {pages} {14} (\bibinfo {year}
  {2015})}\BibitemShut {NoStop}%
\bibitem [{\citenamefont {Maznev}\ \emph {et~al.}(2013)\citenamefont {Maznev},
  \citenamefont {Every},\ and\ \citenamefont
  {Wright}}]{Maznev_WaveMotion50_776_2013}%
  \BibitemOpen
  \bibfield  {author} {\bibinfo {author} {\bibfnamefont {A.}~\bibnamefont
  {Maznev}}, \bibinfo {author} {\bibfnamefont {A.}~\bibnamefont {Every}}, \
  and\ \bibinfo {author} {\bibfnamefont {O.}~\bibnamefont {Wright}},\ }\href
  {\doibase https://doi.org/10.1016/j.wavemoti.2013.02.006} {\bibfield
  {journal} {\bibinfo  {journal} {Wave Motion}\ }\textbf {\bibinfo {volume}
  {50}},\ \bibinfo {pages} {776 } (\bibinfo {year} {2013})}\BibitemShut
  {NoStop}%
\bibitem [{\citenamefont {Liang}\ \emph {et~al.}(2010)\citenamefont {Liang},
  \citenamefont {Guo}, \citenamefont {Tu}, \citenamefont {Zhang},\ and\
  \citenamefont {Cheng}}]{Liang_NatMater9_989_2010}%
  \BibitemOpen
  \bibfield  {author} {\bibinfo {author} {\bibfnamefont {B.}~\bibnamefont
  {Liang}}, \bibinfo {author} {\bibfnamefont {X.~S.}\ \bibnamefont {Guo}},
  \bibinfo {author} {\bibfnamefont {J.}~\bibnamefont {Tu}}, \bibinfo {author}
  {\bibfnamefont {D.}~\bibnamefont {Zhang}}, \ and\ \bibinfo {author}
  {\bibfnamefont {J.~C.}\ \bibnamefont {Cheng}},\ }\href {\doibase
  10.1038/nmat2881} {\bibfield  {journal} {\bibinfo  {journal} {Nat. Mater.}\
  }\textbf {\bibinfo {volume} {9}},\ \bibinfo {pages} {989} (\bibinfo {year}
  {2010})}\BibitemShut {NoStop}%
\bibitem [{\citenamefont {Popa}\ and\ \citenamefont
  {Cummer}(2014)}]{Popa_NatComm5_3398_2014}%
  \BibitemOpen
  \bibfield  {author} {\bibinfo {author} {\bibfnamefont {B.-I.}\ \bibnamefont
  {Popa}}\ and\ \bibinfo {author} {\bibfnamefont {S.~A.}\ \bibnamefont
  {Cummer}},\ }\href {\doibase 10.1038/ncomms4398} {\bibfield  {journal}
  {\bibinfo  {journal} {Nat. Comm.}\ }\textbf {\bibinfo {volume} {5}},\
  \bibinfo {pages} {3398} (\bibinfo {year} {2014})}\BibitemShut {NoStop}%
\bibitem [{\citenamefont {Fleury}\ \emph {et~al.}(2014)\citenamefont {Fleury},
  \citenamefont {Sounas}, \citenamefont {Sieck}, \citenamefont {Haberman},\
  and\ \citenamefont {Al\`u}}]{Fleury_Science343_516_2014}%
  \BibitemOpen
  \bibfield  {author} {\bibinfo {author} {\bibfnamefont {R.}~\bibnamefont
  {Fleury}}, \bibinfo {author} {\bibfnamefont {D.~L.}\ \bibnamefont {Sounas}},
  \bibinfo {author} {\bibfnamefont {C.~F.}\ \bibnamefont {Sieck}}, \bibinfo
  {author} {\bibfnamefont {M.~R.}\ \bibnamefont {Haberman}}, \ and\ \bibinfo
  {author} {\bibfnamefont {A.}~\bibnamefont {Al\`u}},\ }\href@noop {}
  {\bibfield  {journal} {\bibinfo  {journal} {Science}\ }\textbf {\bibinfo
  {volume} {343}},\ \bibinfo {pages} {516} (\bibinfo {year}
  {2014})}\BibitemShut {NoStop}%
\bibitem [{\citenamefont {Khanikaev}\ \emph {et~al.}(2015)\citenamefont
  {Khanikaev}, \citenamefont {Fleury}, \citenamefont {Mousavi},\ and\
  \citenamefont {Al\`u}}]{Khanikaev_NatComm6_8260_2015}%
  \BibitemOpen
  \bibfield  {author} {\bibinfo {author} {\bibfnamefont {A.~B.}\ \bibnamefont
  {Khanikaev}}, \bibinfo {author} {\bibfnamefont {R.}~\bibnamefont {Fleury}},
  \bibinfo {author} {\bibfnamefont {S.~H.}\ \bibnamefont {Mousavi}}, \ and\
  \bibinfo {author} {\bibfnamefont {A.}~\bibnamefont {Al\`u}},\ }\href
  {\doibase 10.1038/ncomms9260} {\bibfield  {journal} {\bibinfo  {journal}
  {Nat. Comm.}\ }\textbf {\bibinfo {volume} {6}},\ \bibinfo {pages} {8260}
  (\bibinfo {year} {2015})}\BibitemShut {NoStop}%
\bibitem [{\citenamefont {Huber}(2016)}]{Huber_NatPhys12_621_2016}%
  \BibitemOpen
  \bibfield  {author} {\bibinfo {author} {\bibfnamefont {S.~D.}\ \bibnamefont
  {Huber}},\ }\href {\doibase 10.1038/nphys3801} {\bibfield  {journal}
  {\bibinfo  {journal} {Nat. Phys.}\ }\textbf {\bibinfo {volume} {12}},\
  \bibinfo {pages} {621} (\bibinfo {year} {2016})}\BibitemShut {NoStop}%
\bibitem [{\citenamefont {Chen}\ \emph {et~al.}(2018)\citenamefont {Chen},
  \citenamefont {Nassar}, \citenamefont {Norris}, \citenamefont {Hu},\ and\
  \citenamefont {Huang}}]{Chen_PRB98_94302_2018}%
  \BibitemOpen
  \bibfield  {author} {\bibinfo {author} {\bibfnamefont {H.}~\bibnamefont
  {Chen}}, \bibinfo {author} {\bibfnamefont {H.}~\bibnamefont {Nassar}},
  \bibinfo {author} {\bibfnamefont {A.~N.}\ \bibnamefont {Norris}}, \bibinfo
  {author} {\bibfnamefont {G.~K.}\ \bibnamefont {Hu}}, \ and\ \bibinfo {author}
  {\bibfnamefont {G.~L.}\ \bibnamefont {Huang}},\ }\href {\doibase
  10.1103/PhysRevB.98.094302} {\bibfield  {journal} {\bibinfo  {journal} {Phys.
  Rev. B}\ }\textbf {\bibinfo {volume} {98}},\ \bibinfo {pages} {094302}
  (\bibinfo {year} {2018})}\BibitemShut {NoStop}%
\bibitem [{\citenamefont {Fleury}\ \emph {et~al.}(2019)\citenamefont {Fleury},
  \citenamefont {Haberman}, \citenamefont {Huang},\ and\ \citenamefont
  {Norris}}]{Fleury_JAcoustSocAm146_719_2019}%
  \BibitemOpen
  \bibfield  {author} {\bibinfo {author} {\bibfnamefont {R.}~\bibnamefont
  {Fleury}}, \bibinfo {author} {\bibfnamefont {M.~R.}\ \bibnamefont
  {Haberman}}, \bibinfo {author} {\bibfnamefont {G.}~\bibnamefont {Huang}}, \
  and\ \bibinfo {author} {\bibfnamefont {A.~N.}\ \bibnamefont {Norris}},\
  }\href {\doibase 10.1121/1.5119133} {\bibfield  {journal} {\bibinfo
  {journal} {J. Acoust. Soc. Am.}\ }\textbf {\bibinfo {volume} {146}},\
  \bibinfo {pages} {719} (\bibinfo {year} {2019})}\BibitemShut {NoStop}%
\bibitem [{\citenamefont {Campbell}(1998)}]{Campbell___98}%
  \BibitemOpen
  \bibfield  {author} {\bibinfo {author} {\bibfnamefont {C.~K.}\ \bibnamefont
  {Campbell}},\ }\href@noop {} {\emph {\bibinfo {title} {Surface Acoustic Wave
  Devices for Mobile and Wireless Communications}}},\ edited by\ \bibinfo
  {editor} {\bibfnamefont {R.}~\bibnamefont {Stern}}\ and\ \bibinfo {editor}
  {\bibfnamefont {M.}~\bibnamefont {Levy}}\ (\bibinfo  {publisher} {Academic
  Press, Inc.},\ \bibinfo {year} {1998})\BibitemShut {NoStop}%
\bibitem [{\citenamefont {Gustafsson}\ \emph {et~al.}(2014)\citenamefont
  {Gustafsson}, \citenamefont {Aref}, \citenamefont {Kockum}, \citenamefont
  {Ekstr\"om}, \citenamefont {Johansson},\ and\ \citenamefont
  {Delsing}}]{Gustafsson_Science346_207_2014}%
  \BibitemOpen
  \bibfield  {author} {\bibinfo {author} {\bibfnamefont {M.~V.}\ \bibnamefont
  {Gustafsson}}, \bibinfo {author} {\bibfnamefont {T.}~\bibnamefont {Aref}},
  \bibinfo {author} {\bibfnamefont {A.~F.}\ \bibnamefont {Kockum}}, \bibinfo
  {author} {\bibfnamefont {M.}~\bibnamefont {Ekstr\"om}}, \bibinfo {author}
  {\bibfnamefont {G.}~\bibnamefont {Johansson}}, \ and\ \bibinfo {author}
  {\bibfnamefont {P.}~\bibnamefont {Delsing}},\ }\href@noop {} {\bibfield
  {journal} {\bibinfo  {journal} {Science}\ }\textbf {\bibinfo {volume}
  {346}},\ \bibinfo {pages} {207} (\bibinfo {year} {2014})}\BibitemShut
  {NoStop}%
\bibitem [{\citenamefont {Golter}\ \emph {et~al.}(2016)\citenamefont {Golter},
  \citenamefont {Oo}, \citenamefont {Amezcua}, \citenamefont {Stewart},\ and\
  \citenamefont {Wang}}]{Golter_PRL116_143602_16}%
  \BibitemOpen
  \bibfield  {author} {\bibinfo {author} {\bibfnamefont {D.~A.}\ \bibnamefont
  {Golter}}, \bibinfo {author} {\bibfnamefont {T.}~\bibnamefont {Oo}}, \bibinfo
  {author} {\bibfnamefont {M.}~\bibnamefont {Amezcua}}, \bibinfo {author}
  {\bibfnamefont {K.~A.}\ \bibnamefont {Stewart}}, \ and\ \bibinfo {author}
  {\bibfnamefont {H.}~\bibnamefont {Wang}},\ }\href {\doibase
  10.1103/PhysRevLett.116.143602} {\bibfield  {journal} {\bibinfo  {journal}
  {Phys. Rev. Lett.}\ }\textbf {\bibinfo {volume} {116}},\ \bibinfo {pages}
  {143602} (\bibinfo {year} {2016})}\BibitemShut {NoStop}%
\bibitem [{\citenamefont {Whiteley}\ \emph {et~al.}(2019)\citenamefont
  {Whiteley}, \citenamefont {Wolfowicz}, \citenamefont {Anderson},
  \citenamefont {Bourassa}, \citenamefont {Ma}, \citenamefont {Ye},
  \citenamefont {Koolstra}, \citenamefont {Satzinger}, \citenamefont {Holt},
  \citenamefont {Heremans}, \citenamefont {Cleland}, \citenamefont {Schuster},
  \citenamefont {Galli},\ and\ \citenamefont
  {Awschalom}}]{Whiteley_NaturePhysics_2019}%
  \BibitemOpen
  \bibfield  {author} {\bibinfo {author} {\bibfnamefont {S.~J.}\ \bibnamefont
  {Whiteley}}, \bibinfo {author} {\bibfnamefont {G.}~\bibnamefont {Wolfowicz}},
  \bibinfo {author} {\bibfnamefont {C.~P.}\ \bibnamefont {Anderson}}, \bibinfo
  {author} {\bibfnamefont {A.}~\bibnamefont {Bourassa}}, \bibinfo {author}
  {\bibfnamefont {H.}~\bibnamefont {Ma}}, \bibinfo {author} {\bibfnamefont
  {M.}~\bibnamefont {Ye}}, \bibinfo {author} {\bibfnamefont {G.}~\bibnamefont
  {Koolstra}}, \bibinfo {author} {\bibfnamefont {K.}~\bibnamefont {Satzinger}},
  \bibinfo {author} {\bibfnamefont {M.~V.}\ \bibnamefont {Holt}}, \bibinfo
  {author} {\bibfnamefont {F.~J.}\ \bibnamefont {Heremans}}, \bibinfo {author}
  {\bibfnamefont {A.~N.}\ \bibnamefont {Cleland}}, \bibinfo {author}
  {\bibfnamefont {D.~I.}\ \bibnamefont {Schuster}}, \bibinfo {author}
  {\bibfnamefont {G.}~\bibnamefont {Galli}}, \ and\ \bibinfo {author}
  {\bibfnamefont {D.~D.}\ \bibnamefont {Awschalom}},\ }\href {\doibase
  10.1038/s41567-019-0420-0} {\bibfield  {journal} {\bibinfo  {journal} {Nat.
  Phys.}\ }\textbf {\bibinfo {volume} {15}},\ \bibinfo {pages} {490} (\bibinfo
  {year} {2019})}\BibitemShut {NoStop}%
\bibitem [{\citenamefont {Lazi{\'{c}}}\ \emph {et~al.}(2019)\citenamefont
  {Lazi{\'{c}}}, \citenamefont {Espinha}, \citenamefont {Yanguas},
  \citenamefont {Gibaja}, \citenamefont {Zamora}, \citenamefont {Ares},
  \citenamefont {Chhowalla}, \citenamefont {Paz}, \citenamefont
  {Palacios~Burgos}, \citenamefont {Hern\'andez-M\'inguez}, \citenamefont
  {Santos},\ and\ \citenamefont {van~der Meulen}}]{Lazic_CommPhys2_113_2019}%
  \BibitemOpen
  \bibfield  {author} {\bibinfo {author} {\bibfnamefont {S.}~\bibnamefont
  {Lazi{\'{c}}}}, \bibinfo {author} {\bibfnamefont {A.}~\bibnamefont
  {Espinha}}, \bibinfo {author} {\bibfnamefont {S.~P.}\ \bibnamefont
  {Yanguas}}, \bibinfo {author} {\bibfnamefont {C.}~\bibnamefont {Gibaja}},
  \bibinfo {author} {\bibfnamefont {F.}~\bibnamefont {Zamora}}, \bibinfo
  {author} {\bibfnamefont {P.}~\bibnamefont {Ares}}, \bibinfo {author}
  {\bibfnamefont {M.}~\bibnamefont {Chhowalla}}, \bibinfo {author}
  {\bibfnamefont {W.~S.}\ \bibnamefont {Paz}}, \bibinfo {author} {\bibfnamefont
  {J.~J.}\ \bibnamefont {Palacios~Burgos}}, \bibinfo {author} {\bibfnamefont
  {A.}~\bibnamefont {Hern\'andez-M\'inguez}}, \bibinfo {author} {\bibfnamefont
  {P.~V.}\ \bibnamefont {Santos}}, \ and\ \bibinfo {author} {\bibfnamefont
  {H.~P.}\ \bibnamefont {van~der Meulen}},\ }\href@noop {} {\bibfield
  {journal} {\bibinfo  {journal} {Comm. Phys.}\ }\textbf {\bibinfo {volume}
  {2}},\ \bibinfo {pages} {113} (\bibinfo {year} {2019})}\BibitemShut {NoStop}%
\bibitem [{\citenamefont {Iikawa}\ \emph {et~al.}(2019)\citenamefont {Iikawa},
  \citenamefont {Hern\'andez-M\'inguez}, \citenamefont {Aharonovich},
  \citenamefont {Nakhaie}, \citenamefont {Liou}, \citenamefont {Lopes},\ and\
  \citenamefont {Santos}}]{Iikawa_APL114_171104_2019}%
  \BibitemOpen
  \bibfield  {author} {\bibinfo {author} {\bibfnamefont {F.}~\bibnamefont
  {Iikawa}}, \bibinfo {author} {\bibfnamefont {A.}~\bibnamefont
  {Hern\'andez-M\'inguez}}, \bibinfo {author} {\bibfnamefont {I.}~\bibnamefont
  {Aharonovich}}, \bibinfo {author} {\bibfnamefont {S.}~\bibnamefont
  {Nakhaie}}, \bibinfo {author} {\bibfnamefont {Y.-T.}\ \bibnamefont {Liou}},
  \bibinfo {author} {\bibfnamefont {J.~M.~J.}\ \bibnamefont {Lopes}}, \ and\
  \bibinfo {author} {\bibfnamefont {P.~V.}\ \bibnamefont {Santos}},\ }\href
  {\doibase 10.1063/1.5093299} {\bibfield  {journal} {\bibinfo  {journal}
  {Appl. Phys. Lett.}\ }\textbf {\bibinfo {volume} {114}},\ \bibinfo {pages}
  {171104} (\bibinfo {year} {2019})}\BibitemShut {NoStop}%
\bibitem [{\citenamefont {Wiele}\ \emph {et~al.}(1998)\citenamefont {Wiele},
  \citenamefont {Haake}, \citenamefont {Rocke},\ and\ \citenamefont
  {Wixforth}}]{Wiele_PRA58_2680_1998}%
  \BibitemOpen
  \bibfield  {author} {\bibinfo {author} {\bibfnamefont {C.}~\bibnamefont
  {Wiele}}, \bibinfo {author} {\bibfnamefont {F.}~\bibnamefont {Haake}},
  \bibinfo {author} {\bibfnamefont {C.}~\bibnamefont {Rocke}}, \ and\ \bibinfo
  {author} {\bibfnamefont {A.}~\bibnamefont {Wixforth}},\ }\href {\doibase
  10.1103/PhysRevA.58.R2680} {\bibfield  {journal} {\bibinfo  {journal} {Phys.
  Rev. A}\ }\textbf {\bibinfo {volume} {58}},\ \bibinfo {pages} {R2680}
  (\bibinfo {year} {1998})}\BibitemShut {NoStop}%
\bibitem [{\citenamefont {Couto}\ \emph {et~al.}(2009)\citenamefont {Couto},
  \citenamefont {Lazi{\'{c}}}, \citenamefont {Iikawa}, \citenamefont {Stotz},
  \citenamefont {Jahn}, \citenamefont {Hey},\ and\ \citenamefont
  {Santos}}]{Couto_NatPhot3_645_2009}%
  \BibitemOpen
  \bibfield  {author} {\bibinfo {author} {\bibfnamefont {O.~D.~D.}\
  \bibnamefont {Couto}}, \bibinfo {author} {\bibfnamefont {S.}~\bibnamefont
  {Lazi{\'{c}}}}, \bibinfo {author} {\bibfnamefont {F.}~\bibnamefont {Iikawa}},
  \bibinfo {author} {\bibfnamefont {J.~A.~H.}\ \bibnamefont {Stotz}}, \bibinfo
  {author} {\bibfnamefont {U.}~\bibnamefont {Jahn}}, \bibinfo {author}
  {\bibfnamefont {R.}~\bibnamefont {Hey}}, \ and\ \bibinfo {author}
  {\bibfnamefont {P.}~\bibnamefont {Santos}},\ }\href@noop {} {\bibfield
  {journal} {\bibinfo  {journal} {Nat. Phot.}\ }\textbf {\bibinfo {volume}
  {3}},\ \bibinfo {pages} {645} (\bibinfo {year} {2009})}\BibitemShut {NoStop}%
\bibitem [{\citenamefont {Gell}\ \emph {et~al.}(2008)\citenamefont {Gell},
  \citenamefont {Ward}, \citenamefont {Young}, \citenamefont {Stevenson},
  \citenamefont {Atkinson}, \citenamefont {Anderson}, \citenamefont {Jones},
  \citenamefont {Ritchie},\ and\ \citenamefont
  {Shields}}]{Gell_APL93_81115_08}%
  \BibitemOpen
  \bibfield  {author} {\bibinfo {author} {\bibfnamefont {J.~R.}\ \bibnamefont
  {Gell}}, \bibinfo {author} {\bibfnamefont {M.~B.}\ \bibnamefont {Ward}},
  \bibinfo {author} {\bibfnamefont {R.~J.}\ \bibnamefont {Young}}, \bibinfo
  {author} {\bibfnamefont {R.~M.}\ \bibnamefont {Stevenson}}, \bibinfo {author}
  {\bibfnamefont {P.}~\bibnamefont {Atkinson}}, \bibinfo {author}
  {\bibfnamefont {D.}~\bibnamefont {Anderson}}, \bibinfo {author}
  {\bibfnamefont {G.~A.~C.}\ \bibnamefont {Jones}}, \bibinfo {author}
  {\bibfnamefont {D.~A.}\ \bibnamefont {Ritchie}}, \ and\ \bibinfo {author}
  {\bibfnamefont {A.~J.}\ \bibnamefont {Shields}},\ }\href {\doibase
  10.1063/1.2976135} {\bibfield  {journal} {\bibinfo  {journal} {Appl. Phys.
  Lett.}\ }\textbf {\bibinfo {volume} {93}},\ \bibinfo {pages} {081115}
  (\bibinfo {year} {2008})}\BibitemShut {NoStop}%
\bibitem [{\citenamefont {Cerda-M\'endez}\ \emph {et~al.}(2010)\citenamefont
  {Cerda-M\'endez}, \citenamefont {Krizhanovskii}, \citenamefont {Wouters},
  \citenamefont {Bradley}, \citenamefont {Biermann}, \citenamefont {Guda},
  \citenamefont {Hey}, \citenamefont {Santos}, \citenamefont {Sarkar},\ and\
  \citenamefont {Skolnick}}]{Cerda-Mendez_PRL105_116402_2010}%
  \BibitemOpen
  \bibfield  {author} {\bibinfo {author} {\bibfnamefont {E.~A.}\ \bibnamefont
  {Cerda-M\'endez}}, \bibinfo {author} {\bibfnamefont {D.~N.}\ \bibnamefont
  {Krizhanovskii}}, \bibinfo {author} {\bibfnamefont {M.}~\bibnamefont
  {Wouters}}, \bibinfo {author} {\bibfnamefont {R.}~\bibnamefont {Bradley}},
  \bibinfo {author} {\bibfnamefont {K.}~\bibnamefont {Biermann}}, \bibinfo
  {author} {\bibfnamefont {K.}~\bibnamefont {Guda}}, \bibinfo {author}
  {\bibfnamefont {R.}~\bibnamefont {Hey}}, \bibinfo {author} {\bibfnamefont
  {P.~V.}\ \bibnamefont {Santos}}, \bibinfo {author} {\bibfnamefont
  {D.}~\bibnamefont {Sarkar}}, \ and\ \bibinfo {author} {\bibfnamefont {M.~S.}\
  \bibnamefont {Skolnick}},\ }\href {\doibase 10.1103/PhysRevLett.105.116402}
  {\bibfield  {journal} {\bibinfo  {journal} {Phys. Rev. Lett.}\ }\textbf
  {\bibinfo {volume} {105}},\ \bibinfo {pages} {116402} (\bibinfo {year}
  {2010})}\BibitemShut {NoStop}%
\bibitem [{\citenamefont {McNeil}\ \emph {et~al.}(2011)\citenamefont {McNeil},
  \citenamefont {Kataoka}, \citenamefont {Ford}, \citenamefont {Barnes},
  \citenamefont {Anderson}, \citenamefont {Jones}, \citenamefont {Farrer},\
  and\ \citenamefont {Ritchie}}]{McNeil_N477_439_11}%
  \BibitemOpen
  \bibfield  {author} {\bibinfo {author} {\bibfnamefont {R.~P.~G.}\
  \bibnamefont {McNeil}}, \bibinfo {author} {\bibfnamefont {M.}~\bibnamefont
  {Kataoka}}, \bibinfo {author} {\bibfnamefont {C.~J.~B.}\ \bibnamefont
  {Ford}}, \bibinfo {author} {\bibfnamefont {C.~H.~W.}\ \bibnamefont {Barnes}},
  \bibinfo {author} {\bibfnamefont {D.}~\bibnamefont {Anderson}}, \bibinfo
  {author} {\bibfnamefont {G.~A.~C.}\ \bibnamefont {Jones}}, \bibinfo {author}
  {\bibfnamefont {I.}~\bibnamefont {Farrer}}, \ and\ \bibinfo {author}
  {\bibfnamefont {D.~A.}\ \bibnamefont {Ritchie}},\ }\href
  {http://dx.doi.org/10.1038/nature10444} {\bibfield  {journal} {\bibinfo
  {journal} {Nature}\ }\textbf {\bibinfo {volume} {477}},\ \bibinfo {pages}
  {439} (\bibinfo {year} {2011})}\BibitemShut {NoStop}%
\bibitem [{\citenamefont {Hermelin}\ \emph {et~al.}(2011)\citenamefont
  {Hermelin}, \citenamefont {Takada}, \citenamefont {Yamamoto}, \citenamefont
  {Tarucha}, \citenamefont {Wieck}, \citenamefont {Saminadayar}, \citenamefont
  {Bauerle},\ and\ \citenamefont {Meunier}}]{Hermelin_N477_435_11}%
  \BibitemOpen
  \bibfield  {author} {\bibinfo {author} {\bibfnamefont {S.}~\bibnamefont
  {Hermelin}}, \bibinfo {author} {\bibfnamefont {S.}~\bibnamefont {Takada}},
  \bibinfo {author} {\bibfnamefont {M.}~\bibnamefont {Yamamoto}}, \bibinfo
  {author} {\bibfnamefont {S.}~\bibnamefont {Tarucha}}, \bibinfo {author}
  {\bibfnamefont {A.~D.}\ \bibnamefont {Wieck}}, \bibinfo {author}
  {\bibfnamefont {L.}~\bibnamefont {Saminadayar}}, \bibinfo {author}
  {\bibfnamefont {C.}~\bibnamefont {Bauerle}}, \ and\ \bibinfo {author}
  {\bibfnamefont {T.}~\bibnamefont {Meunier}},\ }\href
  {http://dx.doi.org/10.1038/nature10416} {\bibfield  {journal} {\bibinfo
  {journal} {Nature}\ }\textbf {\bibinfo {volume} {477}},\ \bibinfo {pages}
  {435} (\bibinfo {year} {2011})}\BibitemShut {NoStop}%
\bibitem [{\citenamefont {Lazi{\'{c}}}\ \emph {et~al.}(2014)\citenamefont
  {Lazi{\'{c}}}, \citenamefont {Violante}, \citenamefont {Cohen}, \citenamefont
  {Hey}, \citenamefont {Rapaport},\ and\ \citenamefont
  {Santos}}]{Lazic_PRB89_85313_14}%
  \BibitemOpen
  \bibfield  {author} {\bibinfo {author} {\bibfnamefont {S.}~\bibnamefont
  {Lazi{\'{c}}}}, \bibinfo {author} {\bibfnamefont {A.}~\bibnamefont
  {Violante}}, \bibinfo {author} {\bibfnamefont {K.}~\bibnamefont {Cohen}},
  \bibinfo {author} {\bibfnamefont {R.}~\bibnamefont {Hey}}, \bibinfo {author}
  {\bibfnamefont {R.}~\bibnamefont {Rapaport}}, \ and\ \bibinfo {author}
  {\bibfnamefont {P.~V.}\ \bibnamefont {Santos}},\ }\href {\doibase
  10.1103/PhysRevB.89.085313} {\bibfield  {journal} {\bibinfo  {journal} {Phys.
  Rev. B}\ }\textbf {\bibinfo {volume} {89}},\ \bibinfo {pages} {085313}
  (\bibinfo {year} {2014})}\BibitemShut {NoStop}%
\bibitem [{\citenamefont {Heil}\ \emph {et~al.}(1982)\citenamefont {Heil},
  \citenamefont {L\"uthi},\ and\ \citenamefont
  {Thalmeier}}]{Heil_PRB25_6515_1982}%
  \BibitemOpen
  \bibfield  {author} {\bibinfo {author} {\bibfnamefont {J.}~\bibnamefont
  {Heil}}, \bibinfo {author} {\bibfnamefont {B.}~\bibnamefont {L\"uthi}}, \
  and\ \bibinfo {author} {\bibfnamefont {P.}~\bibnamefont {Thalmeier}},\ }\href
  {\doibase 10.1103/PhysRevB.25.6515} {\bibfield  {journal} {\bibinfo
  {journal} {Phys. Rev. B}\ }\textbf {\bibinfo {volume} {25}},\ \bibinfo
  {pages} {6515} (\bibinfo {year} {1982})}\BibitemShut {NoStop}%
\bibitem [{\citenamefont {Zhu}\ and\ \citenamefont
  {Rais-Zadeh}(2017)}]{Zhu_IEEEElectrDevLett38_802_2017}%
  \BibitemOpen
  \bibfield  {author} {\bibinfo {author} {\bibfnamefont {H.}~\bibnamefont
  {Zhu}}\ and\ \bibinfo {author} {\bibfnamefont {M.}~\bibnamefont
  {Rais-Zadeh}},\ }\href {\doibase 10.1109/LED.2017.2700013} {\bibfield
  {journal} {\bibinfo  {journal} {IEEE Electr. Dev. Lett.}\ }\textbf {\bibinfo
  {volume} {38}},\ \bibinfo {pages} {802} (\bibinfo {year} {2017})}\BibitemShut
  {NoStop}%
\bibitem [{\citenamefont {Lewis}\ and\ \citenamefont
  {Patterson}(1972)}]{Lewis_APL20_276_1972}%
  \BibitemOpen
  \bibfield  {author} {\bibinfo {author} {\bibfnamefont {M.~F.}\ \bibnamefont
  {Lewis}}\ and\ \bibinfo {author} {\bibfnamefont {E.}~\bibnamefont
  {Patterson}},\ }\href {\doibase 10.1063/1.1654147} {\bibfield  {journal}
  {\bibinfo  {journal} {Appl. Phys. Lett.}\ }\textbf {\bibinfo {volume} {20}},\
  \bibinfo {pages} {276} (\bibinfo {year} {1972})}\BibitemShut {NoStop}%
\bibitem [{\citenamefont {Daniel}(1977)}]{Daniel_JAP48_1732_1977}%
  \BibitemOpen
  \bibfield  {author} {\bibinfo {author} {\bibfnamefont {M.~R.}\ \bibnamefont
  {Daniel}},\ }\href {\doibase 10.1063/1.323815} {\bibfield  {journal}
  {\bibinfo  {journal} {J. Appl. Phys.}\ }\textbf {\bibinfo {volume} {48}},\
  \bibinfo {pages} {1732} (\bibinfo {year} {1977})}\BibitemShut {NoStop}%
\bibitem [{\citenamefont {Weiler}\ \emph {et~al.}(2011)\citenamefont {Weiler},
  \citenamefont {Dreher}, \citenamefont {Heeg}, \citenamefont {Huebl},
  \citenamefont {Gross}, \citenamefont {Brandt},\ and\ \citenamefont
  {Goennenwein}}]{Weiler_PRL106_117601_11}%
  \BibitemOpen
  \bibfield  {author} {\bibinfo {author} {\bibfnamefont {M.}~\bibnamefont
  {Weiler}}, \bibinfo {author} {\bibfnamefont {L.}~\bibnamefont {Dreher}},
  \bibinfo {author} {\bibfnamefont {C.}~\bibnamefont {Heeg}}, \bibinfo {author}
  {\bibfnamefont {H.}~\bibnamefont {Huebl}}, \bibinfo {author} {\bibfnamefont
  {R.}~\bibnamefont {Gross}}, \bibinfo {author} {\bibfnamefont {M.~S.}\
  \bibnamefont {Brandt}}, \ and\ \bibinfo {author} {\bibfnamefont {S.~T.~B.}\
  \bibnamefont {Goennenwein}},\ }\href {\doibase
  10.1103/PhysRevLett.106.117601} {\bibfield  {journal} {\bibinfo  {journal}
  {Phys. Rev. Lett.}\ }\textbf {\bibinfo {volume} {106}},\ \bibinfo {pages}
  {117601} (\bibinfo {year} {2011})}\BibitemShut {NoStop}%
\bibitem [{\citenamefont {Dreher}\ \emph {et~al.}(2012)\citenamefont {Dreher},
  \citenamefont {Weiler}, \citenamefont {Pernpeintner}, \citenamefont {Huebl},
  \citenamefont {Gross}, \citenamefont {Brandt},\ and\ \citenamefont
  {Goennenwein}}]{Dreher_PRB86_134415_12}%
  \BibitemOpen
  \bibfield  {author} {\bibinfo {author} {\bibfnamefont {L.}~\bibnamefont
  {Dreher}}, \bibinfo {author} {\bibfnamefont {M.}~\bibnamefont {Weiler}},
  \bibinfo {author} {\bibfnamefont {M.}~\bibnamefont {Pernpeintner}}, \bibinfo
  {author} {\bibfnamefont {H.}~\bibnamefont {Huebl}}, \bibinfo {author}
  {\bibfnamefont {R.}~\bibnamefont {Gross}}, \bibinfo {author} {\bibfnamefont
  {M.~S.}\ \bibnamefont {Brandt}}, \ and\ \bibinfo {author} {\bibfnamefont
  {S.~T.~B.}\ \bibnamefont {Goennenwein}},\ }\href {\doibase
  10.1103/PhysRevB.86.134415} {\bibfield  {journal} {\bibinfo  {journal} {Phys.
  Rev. B}\ }\textbf {\bibinfo {volume} {86}},\ \bibinfo {pages} {134415}
  (\bibinfo {year} {2012})}\BibitemShut {NoStop}%
\bibitem [{\citenamefont {Sasaki}\ \emph {et~al.}(2017)\citenamefont {Sasaki},
  \citenamefont {Nii}, \citenamefont {Iguchi},\ and\ \citenamefont
  {Onose}}]{Sasaki_PRB95_020407_2017}%
  \BibitemOpen
  \bibfield  {author} {\bibinfo {author} {\bibfnamefont {R.}~\bibnamefont
  {Sasaki}}, \bibinfo {author} {\bibfnamefont {Y.}~\bibnamefont {Nii}},
  \bibinfo {author} {\bibfnamefont {Y.}~\bibnamefont {Iguchi}}, \ and\ \bibinfo
  {author} {\bibfnamefont {Y.}~\bibnamefont {Onose}},\ }\href {\doibase
  10.1103/PhysRevB.95.020407} {\bibfield  {journal} {\bibinfo  {journal} {Phys.
  Rev. B}\ }\textbf {\bibinfo {volume} {95}},\ \bibinfo {pages} {020407(R)}
  (\bibinfo {year} {2017})}\BibitemShut {NoStop}%
\bibitem [{\citenamefont {Kittel}(1958)}]{Kittel_PhysRev110_836_1958}%
  \BibitemOpen
  \bibfield  {author} {\bibinfo {author} {\bibfnamefont {C.}~\bibnamefont
  {Kittel}},\ }\href {\doibase 10.1103/PhysRev.110.836} {\bibfield  {journal}
  {\bibinfo  {journal} {Phys. Rev.}\ }\textbf {\bibinfo {volume} {110}},\
  \bibinfo {pages} {836} (\bibinfo {year} {1958})}\BibitemShut {NoStop}%
\bibitem [{\citenamefont {Matthews}\ and\ \citenamefont {van~de
  Vaart}(1969)}]{Matthews_APL15_373_1969}%
  \BibitemOpen
  \bibfield  {author} {\bibinfo {author} {\bibfnamefont {H.}~\bibnamefont
  {Matthews}}\ and\ \bibinfo {author} {\bibfnamefont {H.}~\bibnamefont {van~de
  Vaart}},\ }\href {\doibase 10.1063/1.1652865} {\bibfield  {journal} {\bibinfo
   {journal} {Appl. Phys. Lett.}\ }\textbf {\bibinfo {volume} {15}},\ \bibinfo
  {pages} {373} (\bibinfo {year} {1969})}\BibitemShut {NoStop}%
\bibitem [{\citenamefont {Tsutsumi}\ \emph {et~al.}(1975)\citenamefont
  {Tsutsumi}, \citenamefont {Bhattacharyya},\ and\ \citenamefont
  {Kumagai}}]{Tsutsumi_JAP46_5072_1975}%
  \BibitemOpen
  \bibfield  {author} {\bibinfo {author} {\bibfnamefont {M.}~\bibnamefont
  {Tsutsumi}}, \bibinfo {author} {\bibfnamefont {T.}~\bibnamefont
  {Bhattacharyya}}, \ and\ \bibinfo {author} {\bibfnamefont {N.}~\bibnamefont
  {Kumagai}},\ }\href {\doibase 10.1063/1.322196} {\bibfield  {journal}
  {\bibinfo  {journal} {J. Appl. Phys.}\ }\textbf {\bibinfo {volume} {46}},\
  \bibinfo {pages} {5072} (\bibinfo {year} {1975})}\BibitemShut {NoStop}%
\bibitem [{\citenamefont {Emtage}(1976)}]{Emtage_PRB13_3063_1976}%
  \BibitemOpen
  \bibfield  {author} {\bibinfo {author} {\bibfnamefont {P.~R.}\ \bibnamefont
  {Emtage}},\ }\href {\doibase 10.1103/PhysRevB.13.3063} {\bibfield  {journal}
  {\bibinfo  {journal} {Phys. Rev. B}\ }\textbf {\bibinfo {volume} {13}},\
  \bibinfo {pages} {3063} (\bibinfo {year} {1976})}\BibitemShut {NoStop}%
\bibitem [{\citenamefont {Komoriya}\ and\ \citenamefont
  {Thomas}(1979)}]{Komoriya_JAP50_6459_1979}%
  \BibitemOpen
  \bibfield  {author} {\bibinfo {author} {\bibfnamefont {G.}~\bibnamefont
  {Komoriya}}\ and\ \bibinfo {author} {\bibfnamefont {G.}~\bibnamefont
  {Thomas}},\ }\href {\doibase 10.1063/1.325740} {\bibfield  {journal}
  {\bibinfo  {journal} {J. Appl. Phys.}\ }\textbf {\bibinfo {volume} {50}},\
  \bibinfo {pages} {6459} (\bibinfo {year} {1979})}\BibitemShut {NoStop}%
\bibitem [{\citenamefont {Camley}(1979)}]{Camley_JAP50_5272_1979}%
  \BibitemOpen
  \bibfield  {author} {\bibinfo {author} {\bibfnamefont {R.~E.}\ \bibnamefont
  {Camley}},\ }\href {\doibase 10.1063/1.326624} {\bibfield  {journal}
  {\bibinfo  {journal} {J. Appl. Phys.}\ }\textbf {\bibinfo {volume} {50}},\
  \bibinfo {pages} {5272} (\bibinfo {year} {1979})}\BibitemShut {NoStop}%
\bibitem [{\citenamefont {Shimizu}\ \emph {et~al.}(1980)\citenamefont
  {Shimizu}, \citenamefont {Hasegawas},\ and\ \citenamefont
  {Yamada}}]{Shimizu_ElectCommJpn63_1_1980}%
  \BibitemOpen
  \bibfield  {author} {\bibinfo {author} {\bibfnamefont {Y.}~\bibnamefont
  {Shimizu}}, \bibinfo {author} {\bibfnamefont {K.}~\bibnamefont {Hasegawas}},
  \ and\ \bibinfo {author} {\bibfnamefont {T.}~\bibnamefont {Yamada}},\ }\href
  {\doibase 10.1002/ecja.4400630302} {\bibfield  {journal} {\bibinfo  {journal}
  {Electron. Commun. Jpn., Part I}\ }\textbf {\bibinfo {volume} {63}},\
  \bibinfo {pages} {1} (\bibinfo {year} {1980})}\BibitemShut {NoStop}%
\bibitem [{\citenamefont {Polzikova}\ \emph {et~al.}(2016)\citenamefont
  {Polzikova}, \citenamefont {Alekseev}, \citenamefont {Pyataikin},
  \citenamefont {Kotelyanskii}, \citenamefont {Luzanov},\ and\ \citenamefont
  {Orlov}}]{Polzikova_AIPAdv6_056306_2016}%
  \BibitemOpen
  \bibfield  {author} {\bibinfo {author} {\bibfnamefont {N.~I.}\ \bibnamefont
  {Polzikova}}, \bibinfo {author} {\bibfnamefont {S.~G.}\ \bibnamefont
  {Alekseev}}, \bibinfo {author} {\bibfnamefont {I.~I.}\ \bibnamefont
  {Pyataikin}}, \bibinfo {author} {\bibfnamefont {I.~M.}\ \bibnamefont
  {Kotelyanskii}}, \bibinfo {author} {\bibfnamefont {V.~A.}\ \bibnamefont
  {Luzanov}}, \ and\ \bibinfo {author} {\bibfnamefont {A.~P.}\ \bibnamefont
  {Orlov}},\ }\href {\doibase 10.1063/1.4943765} {\bibfield  {journal}
  {\bibinfo  {journal} {AIP Advances}\ }\textbf {\bibinfo {volume} {6}},\
  \bibinfo {pages} {056306} (\bibinfo {year} {2016})}\BibitemShut {NoStop}%
\bibitem [{\citenamefont {Ganguly}\ \emph {et~al.}(1976)\citenamefont
  {Ganguly}, \citenamefont {Davis}, \citenamefont {Webb},\ and\ \citenamefont
  {Vittoria}}]{Ganguly_JAP47_2696_1976}%
  \BibitemOpen
  \bibfield  {author} {\bibinfo {author} {\bibfnamefont {A.~K.}\ \bibnamefont
  {Ganguly}}, \bibinfo {author} {\bibfnamefont {K.~L.}\ \bibnamefont {Davis}},
  \bibinfo {author} {\bibfnamefont {D.~C.}\ \bibnamefont {Webb}}, \ and\
  \bibinfo {author} {\bibfnamefont {C.}~\bibnamefont {Vittoria}},\ }\href
  {\doibase 10.1063/1.322991} {\bibfield  {journal} {\bibinfo  {journal} {J.
  Appl. Phys.}\ }\textbf {\bibinfo {volume} {47}},\ \bibinfo {pages} {2696}
  (\bibinfo {year} {1976})}\BibitemShut {NoStop}%
\bibitem [{\citenamefont {Davis}\ \emph {et~al.}(2010)\citenamefont {Davis},
  \citenamefont {Baruth},\ and\ \citenamefont
  {Adenwalla}}]{Davis_APL97_232507_10}%
  \BibitemOpen
  \bibfield  {author} {\bibinfo {author} {\bibfnamefont {S.}~\bibnamefont
  {Davis}}, \bibinfo {author} {\bibfnamefont {A.}~\bibnamefont {Baruth}}, \
  and\ \bibinfo {author} {\bibfnamefont {S.}~\bibnamefont {Adenwalla}},\ }\href
  {\doibase 10.1063/1.3521289} {\bibfield  {journal} {\bibinfo  {journal} {App.
  Phys. Lett.}\ }\textbf {\bibinfo {volume} {97}},\ \bibinfo {pages} {232507}
  (\bibinfo {year} {2010})}\BibitemShut {NoStop}%
\bibitem [{\citenamefont {Gowtham}\ \emph {et~al.}(2015)\citenamefont
  {Gowtham}, \citenamefont {Moriyama}, \citenamefont {Ralph},\ and\
  \citenamefont {Buhrman}}]{Gowtham_JoAP118_233910_15}%
  \BibitemOpen
  \bibfield  {author} {\bibinfo {author} {\bibfnamefont {P.~G.}\ \bibnamefont
  {Gowtham}}, \bibinfo {author} {\bibfnamefont {T.}~\bibnamefont {Moriyama}},
  \bibinfo {author} {\bibfnamefont {D.~C.}\ \bibnamefont {Ralph}}, \ and\
  \bibinfo {author} {\bibfnamefont {R.~A.}\ \bibnamefont {Buhrman}},\ }\href
  {\doibase 10.1063/1.4938390} {\bibfield  {journal} {\bibinfo  {journal} {J.
  Appl. Phys.}\ }\textbf {\bibinfo {volume} {118}},\ \bibinfo {pages} {233910}
  (\bibinfo {year} {2015})}\BibitemShut {NoStop}%
\bibitem [{\citenamefont {Labanowski}\ \emph {et~al.}(2016)\citenamefont
  {Labanowski}, \citenamefont {Jung},\ and\ \citenamefont
  {Salahuddin}}]{Labanowski_APL108_022905_2016}%
  \BibitemOpen
  \bibfield  {author} {\bibinfo {author} {\bibfnamefont {D.}~\bibnamefont
  {Labanowski}}, \bibinfo {author} {\bibfnamefont {A.}~\bibnamefont {Jung}}, \
  and\ \bibinfo {author} {\bibfnamefont {S.}~\bibnamefont {Salahuddin}},\
  }\href {\doibase 10.1063/1.4939914} {\bibfield  {journal} {\bibinfo
  {journal} {Appl. Phys. Lett.}\ }\textbf {\bibinfo {volume} {108}},\ \bibinfo
  {pages} {022905} (\bibinfo {year} {2016})}\BibitemShut {NoStop}%
\bibitem [{\citenamefont {Gowtham}\ \emph {et~al.}(2016)\citenamefont
  {Gowtham}, \citenamefont {Labanowski},\ and\ \citenamefont
  {Salahuddin}}]{Gowtham_PRB94_014436_2016}%
  \BibitemOpen
  \bibfield  {author} {\bibinfo {author} {\bibfnamefont {P.~G.}\ \bibnamefont
  {Gowtham}}, \bibinfo {author} {\bibfnamefont {D.}~\bibnamefont {Labanowski}},
  \ and\ \bibinfo {author} {\bibfnamefont {S.}~\bibnamefont {Salahuddin}},\
  }\href {\doibase 10.1103/PhysRevB.94.014436} {\bibfield  {journal} {\bibinfo
  {journal} {Phys. Rev. B}\ }\textbf {\bibinfo {volume} {94}},\ \bibinfo
  {pages} {014436} (\bibinfo {year} {2016})}\BibitemShut {NoStop}%
\bibitem [{\citenamefont {Li}\ \emph {et~al.}(2017)\citenamefont {Li},
  \citenamefont {Labanowski}, \citenamefont {Salahuddin},\ and\ \citenamefont
  {Lynch}}]{Li_JAP122_043904_2017}%
  \BibitemOpen
  \bibfield  {author} {\bibinfo {author} {\bibfnamefont {X.}~\bibnamefont
  {Li}}, \bibinfo {author} {\bibfnamefont {D.}~\bibnamefont {Labanowski}},
  \bibinfo {author} {\bibfnamefont {S.}~\bibnamefont {Salahuddin}}, \ and\
  \bibinfo {author} {\bibfnamefont {C.~S.}\ \bibnamefont {Lynch}},\ }\href
  {\doibase 10.1063/1.4996102} {\bibfield  {journal} {\bibinfo  {journal} {J.
  Appl. Phys.}\ }\textbf {\bibinfo {volume} {122}},\ \bibinfo {pages} {043904}
  (\bibinfo {year} {2017})}\BibitemShut {NoStop}%
\bibitem [{\citenamefont {Sasaki}\ \emph {et~al.}(2019)\citenamefont {Sasaki},
  \citenamefont {Nii},\ and\ \citenamefont {Onose}}]{Sasaki_PRB99_14418_2019}%
  \BibitemOpen
  \bibfield  {author} {\bibinfo {author} {\bibfnamefont {R.}~\bibnamefont
  {Sasaki}}, \bibinfo {author} {\bibfnamefont {Y.}~\bibnamefont {Nii}}, \ and\
  \bibinfo {author} {\bibfnamefont {Y.}~\bibnamefont {Onose}},\ }\href
  {\doibase 10.1103/PhysRevB.99.014418} {\bibfield  {journal} {\bibinfo
  {journal} {Phys. Rev. B}\ }\textbf {\bibinfo {volume} {99}},\ \bibinfo
  {pages} {014418} (\bibinfo {year} {2019})}\BibitemShut {NoStop}%
\bibitem [{\citenamefont {Thevenard}\ \emph {et~al.}(2013)\citenamefont
  {Thevenard}, \citenamefont {Duquesne}, \citenamefont {Peronne}, \citenamefont
  {von Bardeleben}, \citenamefont {Jaffres}, \citenamefont {Ruttala},
  \citenamefont {George}, \citenamefont {Lema\^{\i}tre},\ and\ \citenamefont
  {Gourdon}}]{Thevenard_PRB87_144402_2013}%
  \BibitemOpen
  \bibfield  {author} {\bibinfo {author} {\bibfnamefont {L.}~\bibnamefont
  {Thevenard}}, \bibinfo {author} {\bibfnamefont {J.-Y.}\ \bibnamefont
  {Duquesne}}, \bibinfo {author} {\bibfnamefont {E.}~\bibnamefont {Peronne}},
  \bibinfo {author} {\bibfnamefont {H.~J.}\ \bibnamefont {von Bardeleben}},
  \bibinfo {author} {\bibfnamefont {H.}~\bibnamefont {Jaffres}}, \bibinfo
  {author} {\bibfnamefont {S.}~\bibnamefont {Ruttala}}, \bibinfo {author}
  {\bibfnamefont {J.-M.}\ \bibnamefont {George}}, \bibinfo {author}
  {\bibfnamefont {A.}~\bibnamefont {Lema\^{\i}tre}}, \ and\ \bibinfo {author}
  {\bibfnamefont {C.}~\bibnamefont {Gourdon}},\ }\href {\doibase
  10.1103/PhysRevB.87.144402} {\bibfield  {journal} {\bibinfo  {journal} {Phys.
  Rev. B}\ }\textbf {\bibinfo {volume} {87}},\ \bibinfo {pages} {144402}
  (\bibinfo {year} {2013})}\BibitemShut {NoStop}%
\bibitem [{\citenamefont {Thevenard}\ \emph {et~al.}(2014)\citenamefont
  {Thevenard}, \citenamefont {Gourdon}, \citenamefont {Prieur}, \citenamefont
  {von Bardeleben}, \citenamefont {Vincent}, \citenamefont {Becerra},
  \citenamefont {Largeau},\ and\ \citenamefont
  {Duquesne}}]{Thevenard_PRB90_094401_2014}%
  \BibitemOpen
  \bibfield  {author} {\bibinfo {author} {\bibfnamefont {L.}~\bibnamefont
  {Thevenard}}, \bibinfo {author} {\bibfnamefont {C.}~\bibnamefont {Gourdon}},
  \bibinfo {author} {\bibfnamefont {J.~Y.}\ \bibnamefont {Prieur}}, \bibinfo
  {author} {\bibfnamefont {H.~J.}\ \bibnamefont {von Bardeleben}}, \bibinfo
  {author} {\bibfnamefont {S.}~\bibnamefont {Vincent}}, \bibinfo {author}
  {\bibfnamefont {L.}~\bibnamefont {Becerra}}, \bibinfo {author} {\bibfnamefont
  {L.}~\bibnamefont {Largeau}}, \ and\ \bibinfo {author} {\bibfnamefont
  {J.-Y.}\ \bibnamefont {Duquesne}},\ }\href {\doibase
  10.1103/PhysRevB.90.094401} {\bibfield  {journal} {\bibinfo  {journal} {Phys.
  Rev. B}\ }\textbf {\bibinfo {volume} {90}},\ \bibinfo {pages} {094401}
  (\bibinfo {year} {2014})}\BibitemShut {NoStop}%
\bibitem [{\citenamefont {Thevenard}\ \emph {et~al.}(2016)\citenamefont
  {Thevenard}, \citenamefont {Camara}, \citenamefont {Majrab}, \citenamefont
  {Bernard}, \citenamefont {Rovillain}, \citenamefont {Lema\^{\i}tre},
  \citenamefont {Gourdon},\ and\ \citenamefont
  {Duquesne}}]{Thevenard_PRB93_134430_16}%
  \BibitemOpen
  \bibfield  {author} {\bibinfo {author} {\bibfnamefont {L.}~\bibnamefont
  {Thevenard}}, \bibinfo {author} {\bibfnamefont {I.~S.}\ \bibnamefont
  {Camara}}, \bibinfo {author} {\bibfnamefont {S.}~\bibnamefont {Majrab}},
  \bibinfo {author} {\bibfnamefont {M.}~\bibnamefont {Bernard}}, \bibinfo
  {author} {\bibfnamefont {P.}~\bibnamefont {Rovillain}}, \bibinfo {author}
  {\bibfnamefont {A.}~\bibnamefont {Lema\^{\i}tre}}, \bibinfo {author}
  {\bibfnamefont {C.}~\bibnamefont {Gourdon}}, \ and\ \bibinfo {author}
  {\bibfnamefont {J.-Y.}\ \bibnamefont {Duquesne}},\ }\href {\doibase
  10.1103/PhysRevB.93.134430} {\bibfield  {journal} {\bibinfo  {journal} {Phys.
  Rev. B}\ }\textbf {\bibinfo {volume} {93}},\ \bibinfo {pages} {134430}
  (\bibinfo {year} {2016})}\BibitemShut {NoStop}%
\bibitem [{\citenamefont {Walowski}\ \emph {et~al.}(2008)\citenamefont
  {Walowski}, \citenamefont {Kaufmann}, \citenamefont {Lenk}, \citenamefont
  {Hamann}, \citenamefont {McCord},\ and\ \citenamefont
  {Münzenberg}}]{Walowski_JPhysD41_164016_2008}%
  \BibitemOpen
  \bibfield  {author} {\bibinfo {author} {\bibfnamefont {J.}~\bibnamefont
  {Walowski}}, \bibinfo {author} {\bibfnamefont {M.~D.}\ \bibnamefont
  {Kaufmann}}, \bibinfo {author} {\bibfnamefont {B.}~\bibnamefont {Lenk}},
  \bibinfo {author} {\bibfnamefont {C.}~\bibnamefont {Hamann}}, \bibinfo
  {author} {\bibfnamefont {J.}~\bibnamefont {McCord}}, \ and\ \bibinfo {author}
  {\bibfnamefont {M.}~\bibnamefont {Münzenberg}},\ }\href {\doibase
  10.1088/0022-3727/41/16/164016} {\bibfield  {journal} {\bibinfo  {journal}
  {J. Phys. D: Appl. Phys.}\ }\textbf {\bibinfo {volume} {41}},\ \bibinfo
  {pages} {164016} (\bibinfo {year} {2008})}\BibitemShut {NoStop}%
\bibitem [{\citenamefont {Herfort}\ \emph {et~al.}(2005)\citenamefont
  {Herfort}, \citenamefont {Sch\"onherr}, \citenamefont {Kawaharazuka},
  \citenamefont {Ramsteiner},\ and\ \citenamefont
  {Ploog}}]{Herfort_JCrysGrowth278_666_2005}%
  \BibitemOpen
  \bibfield  {author} {\bibinfo {author} {\bibfnamefont {J.}~\bibnamefont
  {Herfort}}, \bibinfo {author} {\bibfnamefont {H.-P.}\ \bibnamefont
  {Sch\"onherr}}, \bibinfo {author} {\bibfnamefont {A.}~\bibnamefont
  {Kawaharazuka}}, \bibinfo {author} {\bibfnamefont {M.}~\bibnamefont
  {Ramsteiner}}, \ and\ \bibinfo {author} {\bibfnamefont {K.}~\bibnamefont
  {Ploog}},\ }\href {\doibase https://doi.org/10.1016/j.jcrysgro.2004.12.124}
  {\bibfield  {journal} {\bibinfo  {journal} {J. Cryst. Growth}\ }\textbf
  {\bibinfo {volume} {278}},\ \bibinfo {pages} {666 } (\bibinfo {year}
  {2005})}\BibitemShut {NoStop}%
\bibitem [{\citenamefont {Ionescu}\ \emph {et~al.}(2005)\citenamefont
  {Ionescu}, \citenamefont {Vaz}, \citenamefont {Trypiniotis}, \citenamefont
  {G\"urtler}, \citenamefont {Garc\'{\i}a-Miquel}, \citenamefont {Bland},
  \citenamefont {Vickers}, \citenamefont {Dalgliesh}, \citenamefont
  {Langridge}, \citenamefont {Bugoslavsky}, \citenamefont {Miyoshi},
  \citenamefont {Cohen},\ and\ \citenamefont
  {Ziebeck}}]{Ionescu_PRB71_94401_2005}%
  \BibitemOpen
  \bibfield  {author} {\bibinfo {author} {\bibfnamefont {A.}~\bibnamefont
  {Ionescu}}, \bibinfo {author} {\bibfnamefont {C.~A.~F.}\ \bibnamefont {Vaz}},
  \bibinfo {author} {\bibfnamefont {T.}~\bibnamefont {Trypiniotis}}, \bibinfo
  {author} {\bibfnamefont {C.~M.}\ \bibnamefont {G\"urtler}}, \bibinfo {author}
  {\bibfnamefont {H.}~\bibnamefont {Garc\'{\i}a-Miquel}}, \bibinfo {author}
  {\bibfnamefont {J.~A.~C.}\ \bibnamefont {Bland}}, \bibinfo {author}
  {\bibfnamefont {M.~E.}\ \bibnamefont {Vickers}}, \bibinfo {author}
  {\bibfnamefont {R.~M.}\ \bibnamefont {Dalgliesh}}, \bibinfo {author}
  {\bibfnamefont {S.}~\bibnamefont {Langridge}}, \bibinfo {author}
  {\bibfnamefont {Y.}~\bibnamefont {Bugoslavsky}}, \bibinfo {author}
  {\bibfnamefont {Y.}~\bibnamefont {Miyoshi}}, \bibinfo {author} {\bibfnamefont
  {L.~F.}\ \bibnamefont {Cohen}}, \ and\ \bibinfo {author} {\bibfnamefont
  {K.~R.~A.}\ \bibnamefont {Ziebeck}},\ }\href {\doibase
  10.1103/PhysRevB.71.094401} {\bibfield  {journal} {\bibinfo  {journal} {Phys.
  Rev. B}\ }\textbf {\bibinfo {volume} {71}},\ \bibinfo {pages} {094401}
  (\bibinfo {year} {2005})}\BibitemShut {NoStop}%
\bibitem [{\citenamefont {Herfort}\ \emph {et~al.}(2003)\citenamefont
  {Herfort}, \citenamefont {Sch\"onherr},\ and\ \citenamefont
  {Ploog}}]{Herfort_APL83_3912_2003}%
  \BibitemOpen
  \bibfield  {author} {\bibinfo {author} {\bibfnamefont {J.}~\bibnamefont
  {Herfort}}, \bibinfo {author} {\bibfnamefont {H.-P.}\ \bibnamefont
  {Sch\"onherr}}, \ and\ \bibinfo {author} {\bibfnamefont {K.~H.}\ \bibnamefont
  {Ploog}},\ }\href {\doibase 10.1063/1.1625426} {\bibfield  {journal}
  {\bibinfo  {journal} {Appl. Phys. Lett.}\ }\textbf {\bibinfo {volume} {83}},\
  \bibinfo {pages} {3912} (\bibinfo {year} {2003})}\BibitemShut {NoStop}%
\bibitem [{\citenamefont {Thomas}\ \emph {et~al.}(2009)\citenamefont {Thomas},
  \citenamefont {Schumann}, \citenamefont {Vinzelberg}, \citenamefont
  {Arushanov}, \citenamefont {Engelhard}, \citenamefont {Schmidt},\ and\
  \citenamefont {Gemming}}]{Thomas_Nanotechnology20_235604_2009}%
  \BibitemOpen
  \bibfield  {author} {\bibinfo {author} {\bibfnamefont {J.}~\bibnamefont
  {Thomas}}, \bibinfo {author} {\bibfnamefont {J.}~\bibnamefont {Schumann}},
  \bibinfo {author} {\bibfnamefont {H.}~\bibnamefont {Vinzelberg}}, \bibinfo
  {author} {\bibfnamefont {E.}~\bibnamefont {Arushanov}}, \bibinfo {author}
  {\bibfnamefont {R.}~\bibnamefont {Engelhard}}, \bibinfo {author}
  {\bibfnamefont {O.~G.}\ \bibnamefont {Schmidt}}, \ and\ \bibinfo {author}
  {\bibfnamefont {T.}~\bibnamefont {Gemming}},\ }\href {\doibase
  10.1088/0957-4484/20/23/235604} {\bibfield  {journal} {\bibinfo  {journal}
  {Nanotechnology}\ }\textbf {\bibinfo {volume} {20}},\ \bibinfo {pages}
  {235604} (\bibinfo {year} {2009})}\BibitemShut {NoStop}%
\bibitem [{\citenamefont {Gusenbauer}\ \emph {et~al.}(2011)\citenamefont
  {Gusenbauer}, \citenamefont {Ashraf}, \citenamefont {Stangl}, \citenamefont
  {Hesser}, \citenamefont {Plach}, \citenamefont {Meingast}, \citenamefont
  {Kothleitner},\ and\ \citenamefont {Koch}}]{Gusenbauer_PRB83_35319_2011}%
  \BibitemOpen
  \bibfield  {author} {\bibinfo {author} {\bibfnamefont {C.}~\bibnamefont
  {Gusenbauer}}, \bibinfo {author} {\bibfnamefont {T.}~\bibnamefont {Ashraf}},
  \bibinfo {author} {\bibfnamefont {J.}~\bibnamefont {Stangl}}, \bibinfo
  {author} {\bibfnamefont {G.}~\bibnamefont {Hesser}}, \bibinfo {author}
  {\bibfnamefont {T.}~\bibnamefont {Plach}}, \bibinfo {author} {\bibfnamefont
  {A.}~\bibnamefont {Meingast}}, \bibinfo {author} {\bibfnamefont
  {G.}~\bibnamefont {Kothleitner}}, \ and\ \bibinfo {author} {\bibfnamefont
  {R.}~\bibnamefont {Koch}},\ }\href {\doibase 10.1103/PhysRevB.83.035319}
  {\bibfield  {journal} {\bibinfo  {journal} {Phys. Rev. B}\ }\textbf {\bibinfo
  {volume} {83}},\ \bibinfo {pages} {035319} (\bibinfo {year}
  {2011})}\BibitemShut {NoStop}%
\bibitem [{\citenamefont {Lenz}\ \emph {et~al.}(2005)\citenamefont {Lenz},
  \citenamefont {Kosubek}, \citenamefont {Baberschke}, \citenamefont {Wende},
  \citenamefont {Herfort}, \citenamefont {Sch\"onherr},\ and\ \citenamefont
  {Ploog}}]{Lenz_Phys.Rev.B72_144411_2005}%
  \BibitemOpen
  \bibfield  {author} {\bibinfo {author} {\bibfnamefont {K.}~\bibnamefont
  {Lenz}}, \bibinfo {author} {\bibfnamefont {E.}~\bibnamefont {Kosubek}},
  \bibinfo {author} {\bibfnamefont {K.}~\bibnamefont {Baberschke}}, \bibinfo
  {author} {\bibfnamefont {H.}~\bibnamefont {Wende}}, \bibinfo {author}
  {\bibfnamefont {J.}~\bibnamefont {Herfort}}, \bibinfo {author} {\bibfnamefont
  {H.-P.}\ \bibnamefont {Sch\"onherr}}, \ and\ \bibinfo {author} {\bibfnamefont
  {K.~H.}\ \bibnamefont {Ploog}},\ }\href {\doibase 10.1103/PhysRevB.72.144411}
  {\bibfield  {journal} {\bibinfo  {journal} {Phys. Rev. B}\ }\textbf {\bibinfo
  {volume} {72}},\ \bibinfo {pages} {144411} (\bibinfo {year}
  {2005})}\BibitemShut {NoStop}%
\bibitem [{\citenamefont {Wegscheider}\ \emph {et~al.}(2011)\citenamefont
  {Wegscheider}, \citenamefont {K\"aferb\"ock}, \citenamefont {Gusenbauer},
  \citenamefont {Ashraf}, \citenamefont {Koch},\ and\ \citenamefont
  {Jantsch}}]{Wegscheider_PRB84_054461_2011}%
  \BibitemOpen
  \bibfield  {author} {\bibinfo {author} {\bibfnamefont {M.}~\bibnamefont
  {Wegscheider}}, \bibinfo {author} {\bibfnamefont {G.}~\bibnamefont
  {K\"aferb\"ock}}, \bibinfo {author} {\bibfnamefont {C.}~\bibnamefont
  {Gusenbauer}}, \bibinfo {author} {\bibfnamefont {T.}~\bibnamefont {Ashraf}},
  \bibinfo {author} {\bibfnamefont {R.}~\bibnamefont {Koch}}, \ and\ \bibinfo
  {author} {\bibfnamefont {W.}~\bibnamefont {Jantsch}},\ }\href {\doibase
  10.1103/PhysRevB.84.054461} {\bibfield  {journal} {\bibinfo  {journal} {Phys.
  Rev. B}\ }\textbf {\bibinfo {volume} {84}},\ \bibinfo {pages} {054461}
  (\bibinfo {year} {2011})}\BibitemShut {NoStop}%
\bibitem [{\citenamefont {Lenz}\ \emph {et~al.}(2006)\citenamefont {Lenz},
  \citenamefont {Kosubek}, \citenamefont {Baberschke}, \citenamefont {Herfort},
  \citenamefont {Sch\"onherr},\ and\ \citenamefont
  {Ploog}}]{Lenz_pssc3_122_2006}%
  \BibitemOpen
  \bibfield  {author} {\bibinfo {author} {\bibfnamefont {K.}~\bibnamefont
  {Lenz}}, \bibinfo {author} {\bibfnamefont {E.}~\bibnamefont {Kosubek}},
  \bibinfo {author} {\bibfnamefont {K.}~\bibnamefont {Baberschke}}, \bibinfo
  {author} {\bibfnamefont {J.}~\bibnamefont {Herfort}}, \bibinfo {author}
  {\bibfnamefont {H.-P.}\ \bibnamefont {Sch\"onherr}}, \ and\ \bibinfo {author}
  {\bibfnamefont {K.~H.}\ \bibnamefont {Ploog}},\ }\href {\doibase
  10.1002/pssc.20562426} {\bibfield  {journal} {\bibinfo  {journal} {phys.
  stat. sol. (c)}\ }\textbf {\bibinfo {volume} {3}},\ \bibinfo {pages} {122}
  (\bibinfo {year} {2006})}\BibitemShut {NoStop}%
\bibitem [{\citenamefont {O'Handley}(2000)}]{OHandley_230_2000}%
  \BibitemOpen
  \bibfield  {author} {\bibinfo {author} {\bibfnamefont {R.~C.}\ \bibnamefont
  {O'Handley}},\ }\enquote {\bibinfo {title} {Modern magnetic materials,
  principles and applications},}\ \ (\bibinfo  {publisher} {John Wiler \& Sons,
  Inc.},\ \bibinfo {year} {2000})\ Chap.~\bibinfo {chapter} {7}, p.\ \bibinfo
  {pages} {230}\BibitemShut {NoStop}%
\bibitem [{\citenamefont {Scherbakov}\ \emph {et~al.}(2010)\citenamefont
  {Scherbakov}, \citenamefont {Salasyuk}, \citenamefont {Akimov}, \citenamefont
  {Liu}, \citenamefont {Bombeck}, \citenamefont {Br\"uggemann}, \citenamefont
  {Yakovlev}, \citenamefont {Sapega}, \citenamefont {Furdyna},\ and\
  \citenamefont {Bayer}}]{Scherbakov_PRL105_117204_2010}%
  \BibitemOpen
  \bibfield  {author} {\bibinfo {author} {\bibfnamefont {A.~V.}\ \bibnamefont
  {Scherbakov}}, \bibinfo {author} {\bibfnamefont {A.~S.}\ \bibnamefont
  {Salasyuk}}, \bibinfo {author} {\bibfnamefont {A.~V.}\ \bibnamefont
  {Akimov}}, \bibinfo {author} {\bibfnamefont {X.}~\bibnamefont {Liu}},
  \bibinfo {author} {\bibfnamefont {M.}~\bibnamefont {Bombeck}}, \bibinfo
  {author} {\bibfnamefont {C.}~\bibnamefont {Br\"uggemann}}, \bibinfo {author}
  {\bibfnamefont {D.~R.}\ \bibnamefont {Yakovlev}}, \bibinfo {author}
  {\bibfnamefont {V.~F.}\ \bibnamefont {Sapega}}, \bibinfo {author}
  {\bibfnamefont {J.~K.}\ \bibnamefont {Furdyna}}, \ and\ \bibinfo {author}
  {\bibfnamefont {M.}~\bibnamefont {Bayer}},\ }\href {\doibase
  10.1103/PhysRevLett.105.117204} {\bibfield  {journal} {\bibinfo  {journal}
  {Phys. Rev. Lett.}\ }\textbf {\bibinfo {volume} {105}},\ \bibinfo {pages}
  {117204} (\bibinfo {year} {2010})}\BibitemShut {NoStop}%
\bibitem [{\citenamefont {Bombeck}\ \emph {et~al.}(2012)\citenamefont
  {Bombeck}, \citenamefont {Salasyuk}, \citenamefont {Glavin}, \citenamefont
  {Scherbakov}, \citenamefont {Br\"uggemann}, \citenamefont {Yakovlev},
  \citenamefont {Sapega}, \citenamefont {Liu}, \citenamefont {Furdyna},
  \citenamefont {Akimov},\ and\ \citenamefont
  {Bayer}}]{Bombeck_PRB85_195324_2012}%
  \BibitemOpen
  \bibfield  {author} {\bibinfo {author} {\bibfnamefont {M.}~\bibnamefont
  {Bombeck}}, \bibinfo {author} {\bibfnamefont {A.~S.}\ \bibnamefont
  {Salasyuk}}, \bibinfo {author} {\bibfnamefont {B.~A.}\ \bibnamefont
  {Glavin}}, \bibinfo {author} {\bibfnamefont {A.~V.}\ \bibnamefont
  {Scherbakov}}, \bibinfo {author} {\bibfnamefont {C.}~\bibnamefont
  {Br\"uggemann}}, \bibinfo {author} {\bibfnamefont {D.~R.}\ \bibnamefont
  {Yakovlev}}, \bibinfo {author} {\bibfnamefont {V.~F.}\ \bibnamefont
  {Sapega}}, \bibinfo {author} {\bibfnamefont {X.}~\bibnamefont {Liu}},
  \bibinfo {author} {\bibfnamefont {J.~K.}\ \bibnamefont {Furdyna}}, \bibinfo
  {author} {\bibfnamefont {A.~V.}\ \bibnamefont {Akimov}}, \ and\ \bibinfo
  {author} {\bibfnamefont {M.}~\bibnamefont {Bayer}},\ }\href {\doibase
  10.1103/PhysRevB.85.195324} {\bibfield  {journal} {\bibinfo  {journal} {Phys.
  Rev. B}\ }\textbf {\bibinfo {volume} {85}},\ \bibinfo {pages} {195324}
  (\bibinfo {year} {2012})}\BibitemShut {NoStop}%
\bibitem [{\citenamefont {Nakamura}(1988)}]{Nakamura_26_1988}%
  \BibitemOpen
  \bibfield  {author} {\bibinfo {author} {\bibfnamefont {Y.}~\bibnamefont
  {Nakamura}},\ }\enquote {\bibinfo {title} {Landolt-b\"ornstein, new series
  iii/19c},}\ \ (\bibinfo  {publisher} {Springer, Berlin},\ \bibinfo {year}
  {1988})\ p.~\bibinfo {pages} {26}\BibitemShut {NoStop}%
\bibitem [{\citenamefont {Kubaschewski}(1982)}]{Kubaschewski_1982}%
  \BibitemOpen
  \bibfield  {author} {\bibinfo {author} {\bibfnamefont {O.}~\bibnamefont
  {Kubaschewski}},\ }\href@noop {} {\emph {\bibinfo {title} {Iron-Binary Phase
  Diagrams}}}\ (\bibinfo  {publisher} {Springer, Berlin},\ \bibinfo {year}
  {1982})\BibitemShut {NoStop}%
\bibitem [{\citenamefont {Herfort}\ \emph {et~al.}(2004)\citenamefont
  {Herfort}, \citenamefont {Sch\"onherr}, \citenamefont {Friedland},\ and\
  \citenamefont {Ploog}}]{Herfort_JCSTB22_2073_2004}%
  \BibitemOpen
  \bibfield  {author} {\bibinfo {author} {\bibfnamefont {J.}~\bibnamefont
  {Herfort}}, \bibinfo {author} {\bibfnamefont {H.-P.}\ \bibnamefont
  {Sch\"onherr}}, \bibinfo {author} {\bibfnamefont {K.-J.}\ \bibnamefont
  {Friedland}}, \ and\ \bibinfo {author} {\bibfnamefont {K.~H.}\ \bibnamefont
  {Ploog}},\ }\href {\doibase 10.1116/1.1768528} {\bibfield  {journal}
  {\bibinfo  {journal} {J. Vac. Sci. Technol. B}\ }\textbf {\bibinfo {volume}
  {22}},\ \bibinfo {pages} {2073} (\bibinfo {year} {2004})}\BibitemShut
  {NoStop}%
\bibitem [{\citenamefont {Rayleigh}(1885)}]{Rayleigh_PLMSs1-17_4_85}%
  \BibitemOpen
  \bibfield  {author} {\bibinfo {author} {\bibfnamefont {L.}~\bibnamefont
  {Rayleigh}},\ }\href {\doibase 10.1112/plms/s1-17.1.4} {\bibfield  {journal}
  {\bibinfo  {journal} {Proc. London Math. Soc.}\ }\textbf {\bibinfo {volume}
  {s1-17}},\ \bibinfo {pages} {4} (\bibinfo {year} {1885})}\BibitemShut
  {NoStop}%
\bibitem [{\citenamefont {Herfort}\ \emph {et~al.}(2008)\citenamefont
  {Herfort}, \citenamefont {Sch\"onherr},\ and\ \citenamefont
  {Jenichen}}]{Herfort_JAP103_7_2008}%
  \BibitemOpen
  \bibfield  {author} {\bibinfo {author} {\bibfnamefont {J.}~\bibnamefont
  {Herfort}}, \bibinfo {author} {\bibfnamefont {H.-P.}\ \bibnamefont
  {Sch\"onherr}}, \ and\ \bibinfo {author} {\bibfnamefont {B.}~\bibnamefont
  {Jenichen}},\ }\href {\doibase 10.1063/1.2831333} {\bibfield  {journal}
  {\bibinfo  {journal} {J. Appl. Phys.}\ }\textbf {\bibinfo {volume} {103}},\
  \bibinfo {pages} {07B506} (\bibinfo {year} {2008})}\BibitemShut {NoStop}%
\bibitem [{\citenamefont {Stamps}\ and\ \citenamefont
  {Hillebrands}(1991)}]{Stamps_PRB44_12417_1991}%
  \BibitemOpen
  \bibfield  {author} {\bibinfo {author} {\bibfnamefont {R.~L.}\ \bibnamefont
  {Stamps}}\ and\ \bibinfo {author} {\bibfnamefont {B.}~\bibnamefont
  {Hillebrands}},\ }\href {\doibase 10.1103/PhysRevB.44.12417} {\bibfield
  {journal} {\bibinfo  {journal} {Phys. Rev. B}\ }\textbf {\bibinfo {volume}
  {44}},\ \bibinfo {pages} {12417} (\bibinfo {year} {1991})}\BibitemShut
  {NoStop}%
\bibitem [{\citenamefont {Szymanski}\ \emph {et~al.}(1991)\citenamefont
  {Szymanski}, \citenamefont {Jankowski}, \citenamefont {Dobrzynski},
  \citenamefont {Wisniewski},\ and\ \citenamefont
  {Bednarski}}]{Szymanski_JPCM3_4005_1991}%
  \BibitemOpen
  \bibfield  {author} {\bibinfo {author} {\bibfnamefont {M.}~\bibnamefont
  {Szymanski}}, \bibinfo {author} {\bibfnamefont {M.}~\bibnamefont
  {Jankowski}}, \bibinfo {author} {\bibfnamefont {L.}~\bibnamefont
  {Dobrzynski}}, \bibinfo {author} {\bibfnamefont {A.}~\bibnamefont
  {Wisniewski}}, \ and\ \bibinfo {author} {\bibfnamefont {S.}~\bibnamefont
  {Bednarski}},\ }\href {\doibase 10.1088/0953-8984/3/22/012} {\bibfield
  {journal} {\bibinfo  {journal} {J. Phys. Condens. Matter}\ }\textbf {\bibinfo
  {volume} {3}},\ \bibinfo {pages} {4005} (\bibinfo {year} {1991})}\BibitemShut
  {NoStop}%
\bibitem [{\citenamefont {Gilbert}(1955)}]{Gilbert_PR100_1243_1955}%
  \BibitemOpen
  \bibfield  {author} {\bibinfo {author} {\bibfnamefont {T.~G.}\ \bibnamefont
  {Gilbert}},\ }\href@noop {} {\bibfield  {journal} {\bibinfo  {journal} {Phys.
  Rev.}\ }\textbf {\bibinfo {volume} {100}},\ \bibinfo {pages} {1243} (\bibinfo
  {year} {1955})}\BibitemShut {NoStop}%
\bibitem [{\citenamefont {Gilbert}(2004)}]{Gilbert_IEEETransMagn40_3443_2004}%
  \BibitemOpen
  \bibfield  {author} {\bibinfo {author} {\bibfnamefont {T.~G.}\ \bibnamefont
  {Gilbert}},\ }\href {\doibase 10.1109/TMAG.2004.836740} {\bibfield  {journal}
  {\bibinfo  {journal} {IEEE Transactions on Magnetics}\ }\textbf {\bibinfo
  {volume} {40}},\ \bibinfo {pages} {3443} (\bibinfo {year}
  {2004})}\BibitemShut {NoStop}%
\bibitem [{\citenamefont {Kuszewski}\ \emph
  {et~al.}(2018{\natexlab{a}})\citenamefont {Kuszewski}, \citenamefont
  {Camara}, \citenamefont {Biarrotte}, \citenamefont {Becerra}, \citenamefont
  {von Bardeleben}, \citenamefont {Torres}, \citenamefont {Lema{\^{\i}}tre},
  \citenamefont {Gourdon}, \citenamefont {Duquesne},\ and\ \citenamefont
  {Thevenard}}]{Kuszewski_JPhysCM30_244003_2018}%
  \BibitemOpen
  \bibfield  {author} {\bibinfo {author} {\bibfnamefont {P.}~\bibnamefont
  {Kuszewski}}, \bibinfo {author} {\bibfnamefont {I.~S.}\ \bibnamefont
  {Camara}}, \bibinfo {author} {\bibfnamefont {N.}~\bibnamefont {Biarrotte}},
  \bibinfo {author} {\bibfnamefont {L.}~\bibnamefont {Becerra}}, \bibinfo
  {author} {\bibfnamefont {J.}~\bibnamefont {von Bardeleben}}, \bibinfo
  {author} {\bibfnamefont {W.~S.}\ \bibnamefont {Torres}}, \bibinfo {author}
  {\bibfnamefont {A.}~\bibnamefont {Lema{\^{\i}}tre}}, \bibinfo {author}
  {\bibfnamefont {C.}~\bibnamefont {Gourdon}}, \bibinfo {author} {\bibfnamefont
  {J.-Y.}\ \bibnamefont {Duquesne}}, \ and\ \bibinfo {author} {\bibfnamefont
  {L.}~\bibnamefont {Thevenard}},\ }\href {\doibase 10.1088/1361-648x/aac152}
  {\bibfield  {journal} {\bibinfo  {journal} {J. Phys. Condens. Matter}\
  }\textbf {\bibinfo {volume} {30}},\ \bibinfo {pages} {244003} (\bibinfo
  {year} {2018}{\natexlab{a}})}\BibitemShut {NoStop}%
\bibitem [{\citenamefont {Kuszewski}\ \emph
  {et~al.}(2018{\natexlab{b}})\citenamefont {Kuszewski}, \citenamefont
  {Duquesne}, \citenamefont {Becerra}, \citenamefont {Lema\^{\i}tre},
  \citenamefont {Vincent}, \citenamefont {Majrab}, \citenamefont {Margaillan},
  \citenamefont {Gourdon},\ and\ \citenamefont
  {Thevenard}}]{Kuszewski_PRApp10_034036_2018}%
  \BibitemOpen
  \bibfield  {author} {\bibinfo {author} {\bibfnamefont {P.}~\bibnamefont
  {Kuszewski}}, \bibinfo {author} {\bibfnamefont {J.-Y.}\ \bibnamefont
  {Duquesne}}, \bibinfo {author} {\bibfnamefont {L.}~\bibnamefont {Becerra}},
  \bibinfo {author} {\bibfnamefont {A.}~\bibnamefont {Lema\^{\i}tre}}, \bibinfo
  {author} {\bibfnamefont {S.}~\bibnamefont {Vincent}}, \bibinfo {author}
  {\bibfnamefont {S.}~\bibnamefont {Majrab}}, \bibinfo {author} {\bibfnamefont
  {F.}~\bibnamefont {Margaillan}}, \bibinfo {author} {\bibfnamefont
  {C.}~\bibnamefont {Gourdon}}, \ and\ \bibinfo {author} {\bibfnamefont
  {L.}~\bibnamefont {Thevenard}},\ }\href {\doibase
  10.1103/PhysRevApplied.10.034036} {\bibfield  {journal} {\bibinfo  {journal}
  {Phys. Rev. Appl.}\ }\textbf {\bibinfo {volume} {10}},\ \bibinfo {pages}
  {034036} (\bibinfo {year} {2018}{\natexlab{b}})}\BibitemShut {NoStop}%
\bibitem [{\citenamefont {K\"otter}\ \emph {et~al.}(1989)\citenamefont
  {K\"otter}, \citenamefont {Nembach}, \citenamefont {Wallow},\ and\
  \citenamefont {Nembach}}]{Koetter_MaterSciEngA114_29_1989}%
  \BibitemOpen
  \bibfield  {author} {\bibinfo {author} {\bibfnamefont {G.}~\bibnamefont
  {K\"otter}}, \bibinfo {author} {\bibfnamefont {K.}~\bibnamefont {Nembach}},
  \bibinfo {author} {\bibfnamefont {F.}~\bibnamefont {Wallow}}, \ and\ \bibinfo
  {author} {\bibfnamefont {E.}~\bibnamefont {Nembach}},\ }\href {\doibase
  10.1016/0921-5093(89)90842-3} {\bibfield  {journal} {\bibinfo  {journal}
  {Mater. Sci. Eng., A}\ }\textbf {\bibinfo {volume} {114}},\ \bibinfo {pages}
  {29} (\bibinfo {year} {1989})}\BibitemShut {NoStop}%
\bibitem [{\citenamefont {Auld}(1990)}]{Auld90a}%
  \BibitemOpen
  \bibfield  {author} {\bibinfo {author} {\bibfnamefont {B.~A.}\ \bibnamefont
  {Auld}},\ }\href
  {https://books.google.de/books?id=_2MWAwAAQBAJ&pg=PA145&lpg=PA145&dq=poynting+acoustic&source=bl&ots=sBAUajuKdX&sig=BRL37qLJMEsMT_a21sA5A7GXU9c&hl=en&sa=X&ei=F8CnVL2DDc3Zap_8gqAC#v=onepage&q=piezoelectric&f=false}
  {\emph {\bibinfo {title} {Acoustic Fields and Waves in Solids}}}\ (\bibinfo
  {publisher} {Robert E. Krieger Publishing Company, Inc},\ \bibinfo {address}
  {Malabar, Florida},\ \bibinfo {year} {1990})\BibitemShut {NoStop}%
\bibitem [{\citenamefont {Stancil}\ and\ \citenamefont
  {Prabhakar}(2009)}]{Stancil_2009}%
  \BibitemOpen
  \bibfield  {author} {\bibinfo {author} {\bibfnamefont {D.~A.}\ \bibnamefont
  {Stancil}}\ and\ \bibinfo {author} {\bibfnamefont {A.}~\bibnamefont
  {Prabhakar}},\ }\href@noop {} {\emph {\bibinfo {title} {Spin Waves, Theory
  and Applications}}}\ (\bibinfo  {publisher} {Springer},\ \bibinfo {year}
  {2009})\BibitemShut {NoStop}%
\bibitem [{\citenamefont {Verba}\ \emph {et~al.}(2018)\citenamefont {Verba},
  \citenamefont {Lisenkov}, \citenamefont {Krivorotov}, \citenamefont
  {Tiberkevich},\ and\ \citenamefont {Slavin}}]{Verba_PRApp9_064014_2018}%
  \BibitemOpen
  \bibfield  {author} {\bibinfo {author} {\bibfnamefont {R.}~\bibnamefont
  {Verba}}, \bibinfo {author} {\bibfnamefont {I.}~\bibnamefont {Lisenkov}},
  \bibinfo {author} {\bibfnamefont {I.}~\bibnamefont {Krivorotov}}, \bibinfo
  {author} {\bibfnamefont {V.}~\bibnamefont {Tiberkevich}}, \ and\ \bibinfo
  {author} {\bibfnamefont {A.}~\bibnamefont {Slavin}},\ }\href {\doibase
  10.1103/PhysRevApplied.9.064014} {\bibfield  {journal} {\bibinfo  {journal}
  {Phys. Rev. Appl.}\ }\textbf {\bibinfo {volume} {9}},\ \bibinfo {pages}
  {064014} (\bibinfo {year} {2018})}\BibitemShut {NoStop}%
\bibitem [{\citenamefont {Verba}\ \emph {et~al.}(2019)\citenamefont {Verba},
  \citenamefont {Tiberkevich},\ and\ \citenamefont
  {Slavin}}]{Verba_PRApp12_54061_2019}%
  \BibitemOpen
  \bibfield  {author} {\bibinfo {author} {\bibfnamefont {R.}~\bibnamefont
  {Verba}}, \bibinfo {author} {\bibfnamefont {V.}~\bibnamefont {Tiberkevich}},
  \ and\ \bibinfo {author} {\bibfnamefont {A.}~\bibnamefont {Slavin}},\ }\href
  {\doibase 10.1103/PhysRevApplied.12.054061} {\bibfield  {journal} {\bibinfo
  {journal} {Phys. Rev. Appl.}\ }\textbf {\bibinfo {volume} {12}},\ \bibinfo
  {pages} {054061} (\bibinfo {year} {2019})}\BibitemShut {NoStop}%
\bibitem [{\citenamefont {Jenichen}\ \emph {et~al.}(2015)\citenamefont
  {Jenichen}, \citenamefont {Jahn}, \citenamefont {Nikulin}, \citenamefont
  {Herfort},\ and\ \citenamefont {Kirmse}}]{Jenichen_SST30_114005_2015}%
  \BibitemOpen
  \bibfield  {author} {\bibinfo {author} {\bibfnamefont {B.}~\bibnamefont
  {Jenichen}}, \bibinfo {author} {\bibfnamefont {U.}~\bibnamefont {Jahn}},
  \bibinfo {author} {\bibfnamefont {A.}~\bibnamefont {Nikulin}}, \bibinfo
  {author} {\bibfnamefont {J.}~\bibnamefont {Herfort}}, \ and\ \bibinfo
  {author} {\bibfnamefont {H.}~\bibnamefont {Kirmse}},\ }\href {\doibase
  10.1088/0268-1242/30/11/114005} {\bibfield  {journal} {\bibinfo  {journal}
  {Semicond. Sci. Technol.}\ }\textbf {\bibinfo {volume} {30}},\ \bibinfo
  {pages} {114005} (\bibinfo {year} {2015})}\BibitemShut {NoStop}%
\bibitem [{\citenamefont {Gaucher}\ \emph {et~al.}(2017)\citenamefont
  {Gaucher}, \citenamefont {Jenichen}, \citenamefont {Kalt}, \citenamefont
  {Jahn}, \citenamefont {Trampert},\ and\ \citenamefont
  {Herfort}}]{Gaucher_APL110_102103_2017}%
  \BibitemOpen
  \bibfield  {author} {\bibinfo {author} {\bibfnamefont {S.}~\bibnamefont
  {Gaucher}}, \bibinfo {author} {\bibfnamefont {B.}~\bibnamefont {Jenichen}},
  \bibinfo {author} {\bibfnamefont {J.}~\bibnamefont {Kalt}}, \bibinfo {author}
  {\bibfnamefont {U.}~\bibnamefont {Jahn}}, \bibinfo {author} {\bibfnamefont
  {A.}~\bibnamefont {Trampert}}, \ and\ \bibinfo {author} {\bibfnamefont
  {J.}~\bibnamefont {Herfort}},\ }\href {\doibase 10.1063/1.4977833} {\bibfield
   {journal} {\bibinfo  {journal} {Appl. Phys. Lett.}\ }\textbf {\bibinfo
  {volume} {110}},\ \bibinfo {pages} {102103} (\bibinfo {year}
  {2017})}\BibitemShut {NoStop}%
\bibitem [{\citenamefont {Gaucher}\ \emph {et~al.}(2018)\citenamefont
  {Gaucher}, \citenamefont {Jenichen},\ and\ \citenamefont
  {Herfort}}]{Gaucher_SST33_104005_2018}%
  \BibitemOpen
  \bibfield  {author} {\bibinfo {author} {\bibfnamefont {S.}~\bibnamefont
  {Gaucher}}, \bibinfo {author} {\bibfnamefont {B.}~\bibnamefont {Jenichen}}, \
  and\ \bibinfo {author} {\bibfnamefont {J.}~\bibnamefont {Herfort}},\ }\href
  {\doibase 10.1088/1361-6641/aaddf0} {\bibfield  {journal} {\bibinfo
  {journal} {Semicond. Sci. Technol.}\ }\textbf {\bibinfo {volume} {33}},\
  \bibinfo {pages} {104005} (\bibinfo {year} {2018})}\BibitemShut {NoStop}%
\end{thebibliography}

%

\end{document}